\documentclass[aps,twocolumn,superscriptaddress,english,10pt,nofootinbib]{revtex4}

\usepackage{color}
\usepackage{graphicx}
\usepackage{latexsym}
\usepackage{dcolumn}
\usepackage{bm}

\usepackage{hyperref}
\usepackage{amsmath}
\usepackage{parskip}
\usepackage{adjustbox}
\usepackage{mathtools}
\usepackage{empheq}
\usepackage[normalem]{ulem}
\hypersetup{colorlinks=true, linkcolor=magenta, filecolor=cyan, urlcolor=blue}

\newcommand{\cs}{{\mbox{\tiny CS}}}
\newcommand{\GB}{{\mbox{\tiny GB}}}
\newcommand{\EH}{{\mbox{\tiny EH}}}

\def\@Aboxed#1&#2\ENDDNE{%
  \settowidth\@tempdima{$\displaystyle#1{}$}%
  \addtolength\@tempdima{\fboxsep}%
  \addtolength\@tempdima{\fboxrule}%
  \global\@tempdima=\@tempdima
  \kern\@tempdima
  &
  \kern-\@tempdima
  \fcolorbox{red}{yellow}{$\displaystyle #1#2$}
}

\newlength\dlf

\begin{document}

\title{Gravitational Waves in Chern-Simons-Gauss-Bonnet Gravity}

\author{Tatsuya Daniel}
\email{tatsuya\_daniel@brown.edu}
\affiliation{Brown Theoretical Physics Center, Department of Physics, Brown University, Providence, RI 02912, USA}

\author{Leah Jenks}
\email{ljenks@uchicago.edu}
\affiliation{Kavli Institute for Cosmological Physics, University of Chicago, Chicago, IL 60637, USA}

\author{Stephon Alexander}
\email{stephon\_alexander@brown.edu}
\affiliation{Brown Theoretical Physics Center, Department of Physics, Brown University, Providence, RI 02912, USA}

\begin{abstract}
It is known that the four-dimensional effective field theory arising from heterotic string theory is general relativity with both a Chern-Simons and Gauss-Bonnet term. 
    We study the propagation of gravitational waves in this combination of Chern-Simons and Gauss-Bonnet gravity, both of which have an associated scalar field, the axion and the dilaton respectively, that are kinetically coupled. 
    We review how the combination of dynamical Chern-Simons and Gauss-Bonnet gravities can arise from string theory as corrections to general relativity and show how the gravitational wave waveform is modified in such a theory. We compare our results to a novel framework recently introduced for parametrizing the parity-violating sector (Chern-Simons), and use that to guide our construction of a similar parametrization for the parity-conserving (Gauss-Bonnet) sector. 
    In general, we find that the contributions from the parity-violating and parity-conserving sectors are similar. Moreover, the kinetic coupling between the axion and dilaton introduces an extra contribution to the parity-violating sector of the gravitational waves. Using our parametrization, we are able to comment on initial constraints for the theory parameters, including the time variations of the axion and dilaton. 
\end{abstract}

\date{July 7, 2026}

\maketitle

\section{Introduction}

Einstein's theory of general relativity (GR) has been shown to agree remarkably well with observations \cite{LIGO2019-1, LIGO2019-2, LIGO2021, theligoscientificcollaboration2021tests}. However, theoretical and observational challenges suggest that GR may be modified in the strong field regime \cite{Will2014}. These corrections are generally motivated from a high-energy ultraviolet (UV) theory that, at low energies, leads to corrections to GR in an effective field theory (EFT)\footnote{See e.g. \cite{Alexander2021}.}.

GR has been very well constrained in the weak field regime (see e.g. \cite{Shapiro1990GeneralRA, ReynaudJaekel}), and with the ability to detect gravitational waves (GWs) in the last decade, it has become possible to probe the strong field regime of gravity directly using compact objects, such as black holes and neutron stars \cite{gwtc1, gwtc2, gwtc3, Abbott2017}. Thus, GWs have opened up a new avenue for testing potential modifications of GR.

There are a wide range of modified gravity theories and extensions to GR (see e.g. \cite{Capozziello2011, Faraoni2010} for a review), which may be motivated from the fact that, at high energies, GR is non-renormalizable in a quantum theory of gravity. Modifications to GR have also been proposed as alternatives to inflation, dark matter, and dark energy (e.g., \cite{Faraoni2010, Nojiri2007, Nojiri2017, Odintsov2023}). 
Modified gravity theories are either \emph{parity-conserving}, which remain invariant under a parity transformation, or \emph{parity-violating}, which are not invariant under such a transformation. 

Many well-studied modifications to GR incorporate higher curvature terms. Some well-studied parity-conserving theories are Gauss-Bonnet \cite{Boulware1986, Kanti1996, Torii1997, Alexeyev1997, Kawai1998, Kawai1999} and Starobinsky inflation as a specific type of $f(R)$ gravity \cite{Starobinsky1980, Sotiriou2010}. Examples of parity-violating theories include Chern-Simons gravity \cite{Lue1999, Jackiw2003, Alexander2009, Yunes2009}, parity violating extensions to Teleparallel Gravity \cite{Crisostomi:2017ugk, Conroy2019}, Horava-Lifshitz \cite{Horava2009, Zhu2013}, and ghost-free scalar-tensor gravity \cite{Nishizawa:2018srh}. Recent work has also shown that parity-violating gravitational interactions can be constructed from the Kalb-Ramond field \cite{Manton2024}. 

Two well-studied modified gravity theories are Chern-Simons and Gauss-Bonnet gravity. Chern-Simons gravity can be motivated from the context of particle physics \cite{Alvarez-Gaume1983, Weinberg1996} and leptogenesis \cite{Alexander2006, AlexanderGates2006, Lyth2005}, as well as in other areas such as string theory \cite{Green1984, Green:1984ed, Green1987}, loop quantum gravity \cite{Ashtekar2004, Thiemann2001, Rovelli2004} and effective field theories \cite{Alexander2021, Alexander2022chernsimons}. Furthermore, from a phenomenological perspective, such a theory could give rise to parity violation in the cosmic microwave background (CMB) \cite{QUaD2009, Sorbo2011, Shiraishi2013, Shiraishi2016, Philcox2023} and in the gravitational sector \cite{Lue1999, Jackiw2003, Contaldi2008, AlexanderYunes2018, LoutrelYunes2022}. Notably, parity violation in the gravitational sector can lead to \textit{birefringence} in GW propagation, in which the right- and left- handed polarization modes evolve differently in their amplitude and/or velocity.\footnote{For recent work on how birefringence can lead to condensate-induced inflation in Chern-Simons gravity, see \cite{Dorlis2024}.}

Gauss-Bonnet gravity is another well-motivated modified gravity theory, initially arising from an attempt to generalize GR \cite{Lanczos1932, Lanczos1938, Lovelock1970,  Lovelock1971}; it has also been suggested to arise from string theory \cite{Zwiebach1985, Gross1986, Nepomechie1985, Callan1986, Candelas1985}. Its phenomenological implications have been extensively studied, including its predicted effect on compact objects such as black holes and neutron stars \cite{Moura2006, Guo2008, Maeda2009, Pani2009, Kleihaus2011, Ayzenberg2014, Maselli2015, Kleihaus2016, Kokkotas2017, Konoplya2019}, and its implications for inflation \cite{Kanti2015, Chakraborty2018, Odintsov2018, Yi2018, Kawai2019, Odintsov2019, Rashidi2020, Kawai2021, Kawai2021_2, Kawai2023}. 

One avenue to find modifications to GR that are mathematically well-motivated is string theory, a candidate for a quantum theory of gravity and a unified description of the fundamental forces of Nature \cite{Green1987_vol1, Green1987, Polchinski:1998rq, Polchinski1998, Polchinski1994}. In general, constructions of string theory require more than four dimensions. Upon compactification from a higher-dimensional theory to four dimensions, string theory predicts GR plus perturbative corrections in the string tension $\alpha' = \ell_s^2$ \cite{Callan1985, Gross1986, Hull1987, Metsaev1987}, and some of these corrections are quadratic curvature terms \cite{Zwiebach1985, Deser1986}. In general, corrections to the Einstein-Hilbert action are represented locally by higher derivative additions, and coordinate invariance implies that they must consist of second and higher powers of curvatures, and their derivatives \cite{Deser2002}. 

A natural question to ask then, is what would be precisely the effective four-dimensional gravity theory predicted from string theory. In string theory, the heterotic string is a mixture of the right-moving sector of the superstring and the left-moving sector of the bosonic string. The two sectors need different spacetime dimensions to cancel the anomalies; the matching of dimensions is achieved by compactifying the extra dimensions on a compact manifold. Heterotic string theory (HST) possesses a number of attractive features, including that it is chiral and includes gauge fields \cite{Gross1985, Gross:1985rr}. The two possible gauge groups for the heterotic string are $E_8 \times E_8$ and $SO(32)$ \cite{Gross1985}; furthermore, the superstring's spectrum contains no tachyons and has a graviton \cite{Blumenhagen2013}.

For a long time, it has been theorized that the four-dimensional effective action from HST is captured by GR plus a Gauss-Bonnet term. However, this lacks an axion field; the field strength of the Kalb-Ramond 2-form satisfies the Bianchi identity $dH = \alpha' R \wedge R$, and hence this term cannot be truncated out. Upon compactification to four dimensions, this term results in a correction to GR that can be precisely identified as the Pontryagin term of Chern-Simons gravity, which is typically coupled to an axion field. Thus, for HST, the 4D gravity theory cannot be Gauss-Bonnet or Chern-Simons gravities alone, but rather a combination of the two as corrections to GR, a result that does not depend on the choice of compactification \cite{Cano2021}.

GWs are a powerful probe of modified gravity theories. It is well known that the effects of deviations from GR on GWs can generally be characterized by modifications to the GW amplitude and phase, for example by using the parametrized post-Einsteinian formalism (ppE) \cite{Yunes:2009ke, Mirshekari2012, Yunes2013, Yunes2016, Tahura2018, Tahura2019, Ezquiaga2021, Ezquiaga2022}. 
In this paper, we study Chern-Simons-Gauss-Bonnet (CS-GB) gravity by computing the equations of motion of GW propagation for such a theory, which contains both terms and includes a kinetic coupling between the two associated scalar fields, the axion and the dilaton. We map our analytic expressions to the parity-violating framework put forth in \cite{Jenks2023} and provide an explicit extension to the parity-invariant sector. This extension maps to ppE and provides a framework to explicitly parameterize parity-violating and parity-conserving corrections to GR in GW propagation. From this framework and our mapping of the CS-GB parameters to GW observables, we are able to use the constraints on the GW propagation speed, as well as the coupling constant $\alpha'$, to provide initial constraints on the theory. 

The outline of this paper is as follows: after presenting the basics of CS and GB gravities in Section \ref{sec:dcsgbintro}, we review in Section \ref{sec:2} how both theories can arise from HST by summarizing the stringy derivation from \cite{Cano2021} of the 4D effective action, which showed that the result is a combination of CS and GB gravities. In Section \ref{sec:3} we compute the modified field equations, and in Section \ref{sec:4} we calculate the equations of motion for GWs in an FLRW background. From there, we generalize the parametrization of \cite{Jenks2023} by including the parity-conserving sector, and use the full parametrization to place initial constraints on the CS-GB theory parameters, including the time derivatives of the axion and dilaton, in Section \ref{sec:5}. We briefly discuss other effects, directions for future work, and conclude in Section \ref{sec:discussion}.

Throughout this paper, we use geometric units such that $G = c = 1$, and we assume a $(-,+,+,+)$ metric signature; Greek letters ($\mu$,$\nu$,...) range over all spacetime coordinates, Latin letters (i,j,...) range over spatial indices, and square brackets denote anti-symmetrization over indices.

\section{Basics of Chern-Simons and Gauss-Bonnet Gravities}\label{sec:dcsgbintro}
In this section, we review the basics of Chern-Simons (Sec.~\ref{sec:dcs}) and Gauss-Bonnet (Sec.~\ref{sec:gb}) gravities individually, before turning to the combined theory for the remainder of the paper.

\subsection{Chern-Simons Gravity} \label{sec:dcs}

The CS modification of GR arises in different contexts, including in particle physics \cite{Alvarez-Gaume1983, Alexander2022chernsimons} and in string theory, where it arises from the Green-Schwarz anomaly cancellation mechanism \cite{Green1984, Green:1984ed, Green1987}. In other words, in HST, a quantum effect due to a gauge field induces a CS term in the effective low energy 4D action of GR.\footnote{For more discussion and derivation of this, the reader is referred to the review \cite{Alexander2009}.} 

CS gravity is a 4D deformation of GR that can generally be written as
\begin{align}
    S = S_\EH + S_\cs + S_{\varphi} + S_{\text{mat}}, \label{eq:csaction}
\end{align}
where $S_\EH$ is the usual Einstein-Hilbert action of GR
\begin{align}
    S_\EH = \int d^4x\sqrt{-g}R, \label{eq:hilbert}
\end{align}
the CS term is given by
\begin{align}
    S_\cs = \frac{\alpha}{4\kappa}\int d^4x \varphi ^{*}RR, \label{eq:cs}
\end{align}
with $\kappa = (16\pi)^{-1}$, and $\alpha$ is a coupling parameter. The pseudo-scalar field, $\varphi$, is coupled to the Pontryagin density of the spacetime, which is defined as
\begin{align}
    ^{*}RR = *R^{\mu}_{~\nu}{}^{\rho\sigma}R^{\nu}_{~\mu\rho\sigma}, \label{eq:pontryagin}
\end{align}
where the Hodge dual of the Riemann tensor is
\begin{align}
    *R^{\mu}_{~\nu}{}^{\rho\sigma} = \frac{1}{2}\epsilon^{\rho\sigma\alpha\beta}R^{\mu}_{~\nu\alpha\beta},
\end{align}
with $\epsilon^{\rho\sigma\alpha\beta}$ the antisymmetric Levi-Civita tensor. The scalar field term is
\begin{align}
    S_{\varphi} = -\frac{\beta}{2}\int d^4x\sqrt{-g}[g^{\mu\nu}(\nabla_{\mu}\varphi)(\nabla_{\nu}\varphi) + 2V(\varphi)], \label{eq:scalar}
\end{align}
and lastly we can have an additional matter contribution described by
\begin{align}
    S_{\text{mat}} = \int d^4x\sqrt{-g}\mathcal{L}_{\text{mat}},
\end{align}
where $\mathcal{L}_{\text{mat}}$ is a matter Lagrangian density that does not depend on $\varphi$.

If a nonzero potential $V(\varphi)$ is chosen in Eq.~(\ref{eq:scalar}), then a mass for the scalar field usually has to be generated, which would render the field short-ranged. However, Eq.~(\ref{eq:csaction}) has a shift symmetry, and theories with a shift symmetry do not allow mass terms, hence the field must be long-ranged. Thus, we choose to set $V(\varphi)$ = 0 and neglect the potential term.

The pseudo-scalar $\varphi$ is known as the CS coupling field, which can generically be a function of space and time. If $\varphi = constant$, CS gravity reduces to GR, since the Pontryagin term can be expressed as the divergence of the CS topological current $K_{\mu}$,
\begin{align}
    \nabla_{\mu}K^{\mu} = \frac{1}{2}{}^*RR,
\end{align}
where
\begin{align}
    K^{\mu} = \epsilon^{\mu\nu\rho\sigma}\Gamma^n_{\nu m}\bigg(\partial_{\rho}\Gamma^m_{\sigma n} + \frac{2}{3}\Gamma^m_{\rho l}\Gamma^l_{\sigma n}\bigg),
\end{align}
with $\Gamma$ being the Christoffel connection. Upon integration by parts, $S_\cs$ becomes
\begin{align}
    S_\cs = \alpha(\varphi K^{\mu})\biggr\rvert_{\partial\mathcal{V}} - \frac{\alpha}{2}\int_{\mathcal{V}} d^4x\sqrt{-g}(\nabla_{\mu}\varphi)K^{\mu},
\end{align}
and the first term can be discarded because it is evaluated on the boundary of the manifold, while the second term clearly vanishes if $\varphi$ is constant \cite{Alexander2009}.

The addition of the CS terms modifies the Einstein equations by the addition of the C-tensor, $C_{\mu\nu}$, as
\begin{align}
    G_{\mu\nu} + \frac{\alpha}{\kappa}C_{\mu\nu} = \frac{1}{2\kappa}T_{\mu\nu}. \label{eq:csgeom}
\end{align}
The C-tensor is a 4D generalization of the 3D Cotton-York tensor; it is given by
\begin{align}
    C^{\mu\nu} = (\nabla_{\alpha}\varphi)\epsilon^{\alpha\beta\gamma(\mu}\nabla_{\gamma}R^{\nu)}_{~\beta} + [\nabla_{(\alpha}\nabla_{\beta)}\varphi]~^{*}R^{\beta(\mu\nu)\alpha}.
\end{align}

Furthermore, we get an extra equation of motion for $\varphi$ from the variation of the action:
\begin{align}
    \beta\Box\varphi = -\frac{\alpha}{4}~^{*}RR, \label{eq:cssfeom}
\end{align}
which is the Klein-Gordon equation in the presence of a source term. 

Here we note that there are two formulations of CS gravity (Eq.~(\ref{eq:csaction})): the \emph{non-dynamical} formulation ($\alpha$ arbitrary, $\beta = 0$) and the \emph{dynamical} formulation ($\alpha$ and $\beta$ arbitrary but nonzero). These are two distinct theories, because in the dynamical case the scalar field introduces stress-energy into the field equations, which forces vacuum spacetimes to possess a certain amount of ``scalar hair," a feature which is absent in the non-dynamical formulation. In this paper, we will be considering the dynamical formulation of CS gravity, appropriately called \emph{dynamical Chern-Simons} (dCS) gravity.

\subsection{Gauss-Bonnet Gravity}
\label{sec:gb}
Gauss-Bonnet gravity has been well studied and it has been found to exhibit a rich phenomenology (see e.g. \cite{Sotiriou2013, Sotiriou2014, Doneva2018, Silva2018, Antoniou2017, Cunha2019, Konoplya2019}), from producing viable models of inflation to spontaneous scalarization in compact objects, as well as admitting novel black hole solutions that evade the no-hair theorems \cite{Herdeiro2018}. At the classical level, string theory predicts that Einstein's field equations receive next-to-leading-order corrections that are usually described by higher-order curvature terms in the action. In particular, GB terms occur in HST in the 1-loop effective action of the 4D theory, in the Einstein frame \cite{Zwiebach1985, Gross1986, Nepomechie1985, Callan1986, Candelas1985}. 

The GB action is another deformation of GR that can be written as
\begin{align}
    S = S_\EH + S_\GB, \label{eq:gbaction}
\end{align}
with $S_\EH$ given in Eq.~(\ref{eq:hilbert}) and 
\begin{align}
    S_\GB = \int d^4x\sqrt{-g}\bigg[-\frac{1}{2}\partial_{\alpha}\phi\partial^{\alpha}\phi + \alpha f(\phi)\mathcal{X}_4\bigg],
\end{align}
where $\alpha$ is a coupling constant, $\mathcal{X}_4$ is the 4D GB density,
\begin{align}
    \mathcal{X}_4 = R^2 - 4R_{\mu\nu}R^{\mu\nu} + R_{\mu\nu\rho\sigma}R^{\mu\nu\rho\sigma},
\end{align}
and we have included a kinetic term for the scalar field.\footnote{One can include a potential term V($\phi$) as well; however due to shift symmetry we set $V(\phi) = 0$, just like for CS gravity.}

A standard variation of Eq.~(\ref{eq:gbaction}) yields the field equations \cite{Bryant2021}:
\begin{align}
    \Box\phi &= \alpha f'(\phi)\mathcal{X}_4, \label{eq:gbsfeom} \\
    G_{\mu\nu} &= \frac{1}{2}\partial_{\mu}\phi\partial_{\nu}\phi - \frac{1}{4}g_{\mu\nu}\partial_{\alpha}\phi\partial^{\alpha}\phi - \alpha D_{\mu\nu}^{(\phi)} + 8\pi T_{\mu\nu}, \label{eq:gbgeom}
\end{align}
where $G_{\mu\nu}$ is the Einstein tensor, $T_{\mu\nu}$ is the matter stress-energy tensor, and 
\begin{align}
    D_{\mu\nu}^{(\phi)} = (g_{\mu\rho}g_{\nu\sigma} + g_{\mu\sigma}g_{\nu\rho})\epsilon^{\alpha\sigma\lambda\gamma}\nabla_{\kappa}[^{*}R^{\rho\kappa}_{~\lambda\gamma}\partial_{\alpha}f(\phi)].
\end{align}

For $D = 4$, one can see that, when varying Eq.~(\ref{eq:gbaction}) with respect to the inverse metric, the contributions of the GB density to the field equations vanish identically. However, if there is a dynamical scalar field $\phi$ which is coupled to the GB density, the GB term will have non-vanishing contributions to the field equations, even in four dimensions. This scalar field $\phi$ is conventially referred to as the \emph{dilaton}.

The combination of the Einstein-Hilbert and GB terms in the gravitational action is known as \emph{Einstein-Gauss-Bonnet gravity}, and with the inclusion of the dilaton it is referred to as \emph{Einstein-dilaton-Gauss-Bonnet} (EdGB) gravity, which is what we consider in this paper.

\section{Derivation of 4D Effective String Action}\label{sec:2}

Here we review how CS-GB gravity can arise from HST, a result which was derived in \cite{Cano2021}. We outline the most important steps in this section, with more intermediate steps and explanations provided in Appendix \ref{sec:stringyderivation}.  

Our starting point is the ten-dimensional heterotic superstring effective action at first order in $\alpha'$. We will use the action given by \cite{Bergshoeff1989}, which is obtained upon supersymmetrization of the Lorentz-Chern-Simons terms:
\begin{align}
    \hat{S} &= \frac{g_s^2}{16\pi G_N^{(10)}}\int d^{10}x\sqrt{|\hat{g}|}e^{-2\hat{\phi}}\bigg[\hat{R} - 4(\partial\hat{\phi})^2 + \frac{1}{12}\hat{H}^2 \nonumber \\ &+ \frac{\alpha'}{8}\hat{R}_{(-)\mu\nu ab}\hat{R}_{(-)}^{\mu\nu ab} + \mathcal{O}(\alpha'^3)\bigg], \label{eq:heterotic}
\end{align}
where $\hat{R}_{(-)}$ is the curvature of the torsionful spin connection, $\Omega^a{}_{(-) b}$: 
\begin{align}
    \Omega_{(-)}^a{}_{b} = \omega^a_{~b} - \frac{1}{2}H_{\mu~b}^{~a}dx^{\mu},
\end{align}
with $\omega^a_{~b}$ being the usual spin connection, and $a$ and $b$ are Lorentz indices. $\hat{H}$ is the 3-form field strength associated with the Kalb-Ramond 2-form $\hat{B}$,  
\begin{align}
    \hat{H} = d\hat{B} + \frac{\alpha'}{4}\omega_{(-)}^L,
\end{align}
with $\omega_{(-)}^L$ being the Lorentz-Chern-Simons 3-form of the torsionful spin connection, and all of the gauge fields are already truncated. The asymptotic vacuum expectation value of the dilaton is related to the string coupling constant as $g_s = e^{\langle\hat{\phi}_{\infty}\rangle}$, and $G_N^{(10)} = 8\pi^6g_s^2\ell_s^8$ is the ten-dimensional gravitational constant, with $\ell_s$ being the string tension. 

We want to find the simplest compactification and truncation of this theory down to four dimensions. The minimal consistent truncation possible is a direct product compactification on a six-torus, $\mathcal{M}_4 \times \mathrm{T}^6$, where the metric takes the form
\begin{align}
    d\hat{s}^2 = d\Bar{s}^2 + dz^idz^i,~i = 1,...,6, \label{eq:compactification}
\end{align}
where $d\Bar{s}^2$ is the 4D metric in the Jordan frame, the six-torus is parametrized by the coordinates $z_i \sim z_i + 2\pi\ell_s$, and all the Kaluza-Klein vectors and scalars are taken to be trivial. One can check that this compactification ansatz solves all of the 10D equations of motion once the lower-dimensional ones are satisfied, making this a consistent truncation.

This compactification yields the same theory as Eq.~(\ref{eq:heterotic}), except in four dimensions and with a gravitational constant $G_N^{(4)} = G_N^{(10)} / 2\pi\ell_s^6$. After introducing the Bianchi identity in the action along with a Lagrange multiplier $\varphi$ to promote $H$ to be the dynamical field instead of $B$, we obtain
\begin{align}
    \Bar{S} &= \frac{1}{16\pi G_N^{(4)}}\int d^4x\sqrt{|\Bar{g}|}\bigg\{e^{-2(\hat{\phi} - \hat{\phi}_{\infty})}\bigg[\Bar{R} - 4(\partial\hat\phi)^2 \nonumber \\ &+ \frac{1}{12}H^2\bigg] - \frac{1}{3!}H_{\mu\nu\rho}\epsilon^{\mu\nu\rho\sigma}\partial_{\sigma}\varphi + \frac{\alpha'}{8}\mathcal{L}_{R^2} + \mathcal{O}(\alpha'^3)\bigg\}, \label{eq:4dcompact}
\end{align}
where 
\begin{align}
    \mathcal{L}_{R^2} = e^{-2(\hat{\phi} - \hat{\phi}_{\infty})}\Bar{R}_{(-)\mu\nu\rho\sigma}\Bar{R}_{(-)}^{\mu\nu\rho\sigma} - \varphi\Bar{R}_{(-)\mu\nu\rho\sigma}\Tilde{\Bar{R}}_{(-)}^{{\mu\nu\rho\sigma}}.
\end{align}
We can vary Eq.~(\ref{eq:4dcompact}) in terms of $H$ to get a relation between $H$ and $\varphi$, thus allowing us to remove $H$ from the action. The variation of Eq.~(\ref{eq:4dcompact}) with respect to $H$ can be solved by doing an expansion in $\alpha'$, i.e. $H = H^{(0)} + \alpha'H^{(1)} + \alpha'^2H^{(2)} + ...$ . After doing so and plugging $H(\varphi)$ back into the action, we find that Eq.~(\ref{eq:4dcompact}) can be written as\footnote{More details provided in Appendix \ref{sec:stringyderivation}.}
\begin{align}
    \Bar{S} &= \frac{1}{16\pi G_N^{(4)}}\int d^4x\sqrt{|g|}\bigg\{e^{-2(\hat{\phi}-\hat{\phi}_{\infty})}\bigg[\Bar{R} - 4(\partial\hat{\phi})^2\bigg] \nonumber \\ &+ \frac{1}{2}e^{2(\hat{\phi}-\hat{\phi}_{\infty})}(\partial\varphi)^2 + \frac{\alpha'}{8}\mathcal{L}_{R^2}\biggr\rvert_{H^{(0)}} + \mathcal{O}(\alpha'^2)\bigg\}\label{eq:jordanaction},
\end{align}
where
\begin{align}
    \mathcal{L}_{R^2}\biggr\rvert_{H^{(0)}} &= e^{-2(\hat{\phi}-\hat{\phi}_{\infty})}\bigg[\Bar{R}_{\mu\nu\rho\sigma}\Bar{R}^{\mu\nu\rho\sigma} + 6\Bar{G}_{\mu\nu}A^{\mu}A^{\nu} + \frac{7}{4}A^4 \nonumber \\ &- 2\Bar{\nabla}_{\mu}A_{\nu}\Bar{\nabla}^{\mu}A^{\nu} - (\Bar{\nabla}_{\mu}A^{\mu})^2\bigg] - \varphi\Bar{R}_{\mu\nu\rho\sigma}\Tilde{\Bar{R}}^{\mu\nu\rho\sigma} \nonumber \\ &+ \text{total derivatives}, \label{eq:4derivativeterm}
\end{align}
where $A_{\mu} = e^{2(\hat{\phi} - \hat{\phi}_{\infty})}\partial_{\mu}\varphi$ and $\Bar{G}_{\mu\nu}$ is the Einstein tensor.

At this point, we note that Eq.~(\ref{eq:jordanaction}) is in the so-called Jordan frame, or equivalently the string frame. To transform it into the Einstein frame, we need to rescale the metric:
\begin{align}
    \Bar{g}_{\mu\nu} = e^{2(\hat{\phi} - \hat{\phi}_{\infty})}g_{\mu\nu}. \label{eq:conformalrescaling}
\end{align}
Eq.~(\ref{eq:jordanaction}) then becomes
\begin{align}
    S &= \frac{1}{16\pi}\int d^4x\sqrt{|g|}\bigg[R + \frac{1}{2}(\partial\phi)^2 + \frac{1}{2}e^{2\phi}(\partial\varphi)^2 \nonumber \\ &+ \frac{\alpha'}{8}\bigg(e^{-\phi}\mathcal{X}_4 - \varphi R_{\mu\nu\rho\sigma}\Tilde{R}^{\mu\nu\rho\sigma}\bigg) + \mathcal{O}(\alpha'^2)\bigg], \label{eq:action}
\end{align}
where we have set $G_N  = 1$, $\mathcal{X}_4 = R^2 - 4R_{\mu\nu}R^{\mu\nu} + R_{\mu\nu\rho\sigma}R^{\mu\nu\rho\sigma}$ is the 4D GB density, and we have introduced the 4D dilaton $\phi = 2(\hat{\phi} - \hat{\phi}_{\infty})$. We see in Eq.~(\ref{eq:action}) that GB and CS gravities are corrections to GR at linear order in $\alpha'$, with a kinetic coupling between the two scalar fields; we will call $\varphi$ the axion, with $\phi$ being the aforementioned dilaton.

A comment on this stringy derivation is in order, particularly regarding our choice of compactification in Eq.~(\ref{eq:compactification}). The multitude of possible compactification choices, together with a plethora of massless 4D moduli fields originating from the deformation modes of the extra dimensions, leads to vacuum degeneracy and moduli problems \cite{Cicoli2013}. There has been recent progress in achieving moduli stabilization (see e.g. \cite{Anguelova2010, Gukov2004, Cicoli2013, Baumann2014, Bernardo2022} and references therein, as well as \cite{Brandenberger2023, Mcallister2023} for recent reviews); while these results need to be combined with viable string constructions of particle physics, the emergence of the CS and GB terms as corrections to GR in a low-energy EFT is a general prediction of string theory \cite{Zwiebach1985, Alexander2009}.

It is also worth noting that it is rather non-trivial that the only higher derivative corrections of Eq.~(\ref{eq:action}) are the CS and GB terms; there are in principle higher derivative terms that could be present in the action. However, Eq.~(\ref{eq:action}) is a general result, and these terms are not neglected by assuming that the scalar fields are of order $\alpha'$; these terms are just simply not present \cite{Cano2021}.

\section{Field Equations}\label{sec:3}
We find the field equations for CS-GB gravity by  varying Eq.~(\ref{eq:action}) with respect to the dilaton, axion, and inverse metric, respectively, which yields:
\begin{align}
&\nabla^2\phi = e^{2\phi}(\partial\varphi)^2 - \frac{\alpha'}{8}e^{-\phi}\mathcal{X}_4, \label{eq:1} \\
&\nabla_{\mu}(e^{2\phi}\nabla^{\mu}\varphi) = -\frac{\alpha'}{8}R_{\mu\nu\rho\sigma}\Tilde{R}^{\mu\nu\rho\sigma}, \label{eq:2} \\
&G_{\mu\nu} + \frac{\alpha'}{8}\bigg(D_{\mu\nu}^{(\phi)} + 2C_{\mu\nu}\bigg) = 8\pi\bigg(T_{\mu\nu}^{(\phi)} + T_{\mu\nu}^{(\varphi)}\bigg), \label{eq:3}
\end{align}
where
\begin{align}
    D_{\mu\nu}^{(\phi)} &= (g_{\mu\rho}g_{\nu\sigma} + g_{\mu\sigma}g_{\nu\rho})\epsilon^{0\sigma\lambda\gamma}\nabla_{\kappa}[^{*}R^{\rho\kappa}_{~~\lambda\gamma}(e^{-\phi})'], \\
    C^{\mu\nu} &= (\nabla_{\alpha}\varphi)\epsilon^{\alpha\beta\gamma(\mu}\nabla_{\gamma}R^{\nu)}_{~\beta} + [\nabla_{(\alpha}\nabla_{\beta)}\varphi]^*R^{\beta(\mu\nu)\alpha}, \\
    T_{\mu\nu}^{(\phi)} &= \nabla_{\mu}\phi\nabla_{\nu}\phi - \frac{1}{2}g_{\mu\nu}\bigg(\nabla_{\alpha}\phi\nabla^{\alpha}\phi\bigg), \\
    T_{\mu\nu}^{(\varphi)} &= e^{2\phi}\nabla_{\mu}\varphi\nabla_{\nu}\varphi - \frac{1}{2}g_{\mu\nu}e^{2\phi}\nabla_{\alpha}\varphi\nabla^{\alpha}\varphi.
\end{align}

We see that Eqs.~(\ref{eq:1})-(\ref{eq:3}) are a combination of the CS and GB field equations in Section \ref{sec:dcsgbintro}, as expected since in our theory, the CS and GB terms appear as a linear combination at first order in $\alpha'$, in addition to the kinetic coupling between the dilaton and the axion. $D_{\mu\nu}^{(\phi)}$ comes from the GB term, and $C^{\mu\nu}$ is the C-tensor that was introduced in Sec.~\ref{sec:dcs}.

Eq.~(\ref{eq:1}) tells us that the dilaton is sourced by the GB term and the axidilaton coupling, in Eq.~(\ref{eq:2}) the Pontryagin term sources the axidilaton kinetic coupling, and in Eq.~(\ref{eq:3}) a linear combination of CS and GB modifies the GR gravitational field equations.

\section{GWs in FLRW Background}\label{sec:4}
Having derived the equations of motion in the previous section, we now study how the propagation of GWs on a cosmological background is modified from GR in CS-GB gravity. We consider the tensor perturbation
\begin{align}
    ds^2 = g_{\mu\nu}dx^{\mu}dx^{\nu} = a^2(\eta)[-d\eta^2 + (\delta_{ij} + h_{ij})dx^idx^j], \label{eq:pert}
\end{align}
where $h_{ij}$ satisfies the transverse-traceless conditions $\partial^jh_{ij} = h_{ii} = 0$.

Perturbing Eq.~(\ref{eq:3}) using Eq.~(\ref{eq:pert}), and expanding in $\alpha'$ as well as the scalar fields, we obtain the linearized equations
\begin{align}
    \bigg(&1 - \frac{\alpha'}{2a^2}\phi''\bigg) \Box h^j_{~i} + \frac{\alpha'}{2a^2}\epsilon^{pjk}\bigg[\bigg(\varphi'' - 2\mathcal{H}\varphi'\bigg)\partial_ph_{ki}' \nonumber \\ +~ &\varphi'\partial_p\Box h_{ki}\bigg] = 0, \label{eq:hij}
\end{align}
where primes denote derivatives with respect to conformal time. Here we take the probe limit of $\phi$ and $\varphi$, assuming that the effects are small enough such that there is no backreaction onto the metric.

We can write Eq.~(\ref{eq:hij}) in terms of the right- and left-circular basis (R/L) of the two helicity-2 polarizations of GWs:
\begin{align}
    h_{ij} = \begin{pmatrix} \frac{1}{\sqrt{2}}(h_L + h_R) & -\frac{i}{\sqrt{2}}(h_L - h_R) & 0 \\ -\frac{i}{\sqrt{2}}(h_L - h_R) & -\frac{1}{\sqrt{2}}(h_L + h_R) & 0 \\ 0 & 0 & 0 &\end{pmatrix}. \label{eq:gwcircularbasis}
\end{align}

Using Eq.~(\ref{eq:gwcircularbasis}), we find that Eq.~(\ref{eq:hij}) can be written as
\begin{align}
     Ah_{R,L}'' + Bh_{R,L}' + Ch_{R,L} = 0, \label{eq:gw}
\end{align}
where
\begin{align}
    A &= 1 - \frac{\alpha'}{4a^2}\phi'' - \lambda_{R,L}\frac{\alpha'a^2}{2}k\varphi', \\
    B &= 2\mathcal{H} + \frac{\alpha'}{2a^2}\mathcal{H}\phi'' - \lambda_{R,L}\alpha'k\bigg(a^2\mathcal{H}\varphi' + \frac{\varphi''}{a^2}\bigg), \\
    C &= k^2 - 2\mathcal{H}^2 + 6\mathcal{H}' + \frac{\alpha'}{2a^2}\phi''\bigg(4\mathcal{H}^2 - 12\mathcal{H}' - k^2\bigg) \nonumber \\ &- \lambda_{R,L}\frac{\alpha'}{2}k\bigg[a^2\varphi'(4\mathcal{H}^2 + k^2) + \frac{2\mathcal{H}\varphi''}{a^4}\bigg],
\end{align}
and $\lambda_{R,L} = \pm 1$. Since the right-hand side of Eq.~(\ref{eq:gw}) is 0, we can divide by $A$ to rewrite it as 

\begin{align}
    h_{R,L}'' + \Bar{B}h_{R,L}' + \Bar{C}h_{R,L} = 0, \label{eq:hrl}
\end{align}
where $\Bar{B} \equiv B/A$ and $\Bar{C} \equiv C/A$.

Taylor expanding $\Bar{B}$ and $\Bar{C}$ to linear order in $\alpha'$ (see Appendix \ref{sec:gwcoefftaylor} for the explicit form of the expansions), and using the evolution of the background scalar fields,
\begin{align}
    \phi'' &= 2\mathcal{H}\phi' + 2\varphi'^2, \label{eq:phibkgd} \\
    \varphi'' &= 2\mathcal{H}\varphi' - 2\phi'\varphi', \label{eq:varphibkgd}
\end{align}
we have 
\begin{align}
    \Bar{B} &= 2\mathcal{H} - 2\lambda_{R,L}\alpha'k\bigg(\frac{\mathcal{H}}{a}\varphi' - \phi'\varphi'\bigg), \label{eq:b} \\
    \Bar{C} &= k^2\bigg[1 - \frac{\alpha'}{2}\bigg(\frac{\mathcal{H}}{a}\phi' + \varphi'^2\bigg)\bigg]. \label{eq:c}
\end{align}

 Moreover, we have assumed that $k \gg \mathcal{H}$ (i.e. that GW wavelengths are short compared to the expansion of the universe), and $\phi'' \ll (\phi')^2$. Note that we are keeping terms quadratic in the scalar fields, even though we are treating them to be small.

For $\Bar{B}$, we see that we have two overall terms; the first is the background, and the second is the parity-violating modification.\footnote{There is a parity-conserving modification to $\bar{B}$, but those terms are highly suppressed, of order $\mathcal{H}^2$ and $\mathcal{H}\varphi'^2$.} Notice that in CS gravity alone, only the term proportional to $\mathcal{H}\varphi'$ is present, while in GB alone there is no correction to $\bar{B}$. In $\bar{C}$, we see that we only have parity-invariant corrections. We note that the axidilaton coupling actually shows up in $\Bar{C}$ as $e^{2\phi}\varphi'^2$, but in our expansion of the scalar fields, the coupling terms show up beginning at third order in the fields, so the leading order contribution is simply $\varphi'^2$. Similarly here, in CS gravity alone, there is no correction to $\bar{C}$, while in GB gravity one obtains only the first correction proportional to $\mathcal{H}\phi'$.

From the propagation equations, we can find the explicit corrections to $h_{R,L}$. The linear perturbations of GWs can be expressed in spatial Fourier space as
\begin{align}
    h_{R,L}(\eta) = \mathcal{A}_{R,L}(\eta)e^{-i[\theta(\eta) - k_ix^i]}. \label{eq:fourierdecomp}
\end{align}

Plugging Eq.~(\ref{eq:fourierdecomp}) into the equations of motion Eq.~(\ref{eq:hrl}), we find the modified dispersion relation
\begin{align}
    &i\theta'' + \theta'^2 + i\theta'\bigg[2\mathcal{H} - 2\lambda_{R,L}\alpha'k\bigg(\frac{\mathcal{H}}{a}\varphi' - \phi'\varphi'\bigg)\bigg] \nonumber \\ - &k^2\bigg[1 - \frac{\alpha'}{2}\bigg(\frac{1}{a}\mathcal{H}\phi' + \varphi'^2\bigg)\bigg] = 0. \label{eq:moddispersion}
\end{align}
From here, we can linearize the equations of motion by taking $\theta \rightarrow \bar{\theta} + \delta\theta$, where the background $\theta$ is the usual GR solution, $\theta' = k - i\mathcal{H}$. Applying this to Eq.~(\ref{eq:moddispersion}), and performing a series expansion assuming that $\delta\theta \ll \Bar{\theta}$, $\theta'' \ll (\theta')^2$ and $\delta\theta'' \ll \Bar{\theta}\delta\theta'$ \cite{Yunes2010}, we get that
\begin{align}
    \delta\theta' &= i\lambda_{R,L}\alpha'k\bigg(\frac{\mathcal{H}}{a}\varphi' - \phi'\varphi'\bigg) - \frac{\alpha'k}{4}\bigg(\frac{1}{a}\mathcal{H}\phi' + \varphi'^2\bigg). \label{eq:deltatheta}
\end{align}
We see that $\delta\theta$ has both real and imaginary parts, which are associated with velocity birefringence and amplitude birefringence terms, respectively. We can write Eq.~(\ref{eq:deltatheta}) accordingly as
\begin{align}
    \delta\theta = -i\lambda_{R,L}\delta\theta_A + \delta\theta_V, \label{eq:birefringence}
\end{align}
where
\begin{align}
    \delta\theta'_A &= -k\alpha'\bigg(\frac{\mathcal{H}}{a}\varphi' - \phi'\varphi'\bigg), \label{eq:ampbirefringence} \\
    \delta\theta'_V &= -\frac{\alpha'k}{4}\bigg(\frac{1}{a}\mathcal{H}\phi' + \varphi'^2\bigg), \label{eq:velbirefringence}
\end{align}
and the $A$ and $V$ subscripts denote the amplitude and velocity contributions, respectively. 

To simplify Eqs.~(\ref{eq:ampbirefringence}) and (\ref{eq:velbirefringence}) further, we will assume that $\phi'$ and $\varphi'$ vary slowly with respect to the expansion of the universe, and can thus be well approximated by their current values via a Taylor expansion, e.g. $\phi' \approx \phi'_0.$ Furthermore, we will use that $dt = -dz/[H(z)(1+z)]$, as well as the fact that $k$ is a constant in conformal time. Using all of this, we can write the integrals of Eqs.~(\ref{eq:ampbirefringence}) and (\ref{eq:velbirefringence}) as
\begin{align}
    \delta\theta_A &= -k\alpha'\bigg(\varphi'_0\int dz - \phi'_0\varphi'_0\int\frac{dz}{H}\bigg), \label{eq:125} \\
    \delta\theta_V &= -\frac{\alpha'k}{4}\bigg(\phi'_0\int dz + \varphi'_0\int \frac{dz}{H}\bigg). \label{eq:126}
\end{align}
We can now define an effective distance, $D_{\alpha}$ as in \cite{Mirshekari2012, Ezquiaga2022},
\begin{align}
    D_{\alpha} = (1+z)^{1-\alpha}\int\frac{(1+z)^{\alpha - 2}}{H(z)}dz,
\end{align}
as well as an effective redshift parameter, $z_{\alpha}$, such that \cite{Jenks2023}
\begin{align}
    z_{\alpha} = (1+z)^{-\alpha}\int\frac{dz}{(1+z)^{1-\alpha}}.
\end{align}
We note that $D_1 = D_T$, where $D_T$ is the look-back distance, and $D_2 = (1+z)^{-1}D_C = D_A$, where $D_C$ and $D_A$ are the comoving and angular-diameter distances, respectively. We can see that $z_0 = \text{ln}(1+z)$ and $z_1 = z(1+z)^{-1}$.

With these definitions, we can write Eqs.~(\ref{eq:125}) and (\ref{eq:126}) as
\begin{align}
    \delta\theta_A &= \alpha'k(1+z)(D_2\phi'_0\varphi'_0 - z_1\varphi'_0), \\
    \delta\theta_V &= -\frac{\alpha'k(1+z)}{4}\bigg(D_2\varphi'_{0}{}^2 + z_1\phi'_0\bigg),
\end{align}
and we have
\begin{align}
    h_{R,L} = \Bar{h}_{R,L}&\text{exp}\bigg[\mp\alpha'k(1+z)(D_2\phi_0'\varphi'_0 - z_1\varphi'_0)\bigg] \nonumber \\ \times~&\text{exp}\bigg[-\frac{i\alpha'k(1+z)}{4}\bigg(z_1\phi'_0 + D_2\varphi'_0{}^2\bigg)\bigg],
\end{align}
where $\Bar{h}_{R,L}$ is the usual GR expression for the right and left-handed modes. 

\begin{figure*}[htb!]
    \includegraphics[width=0.482\textwidth]{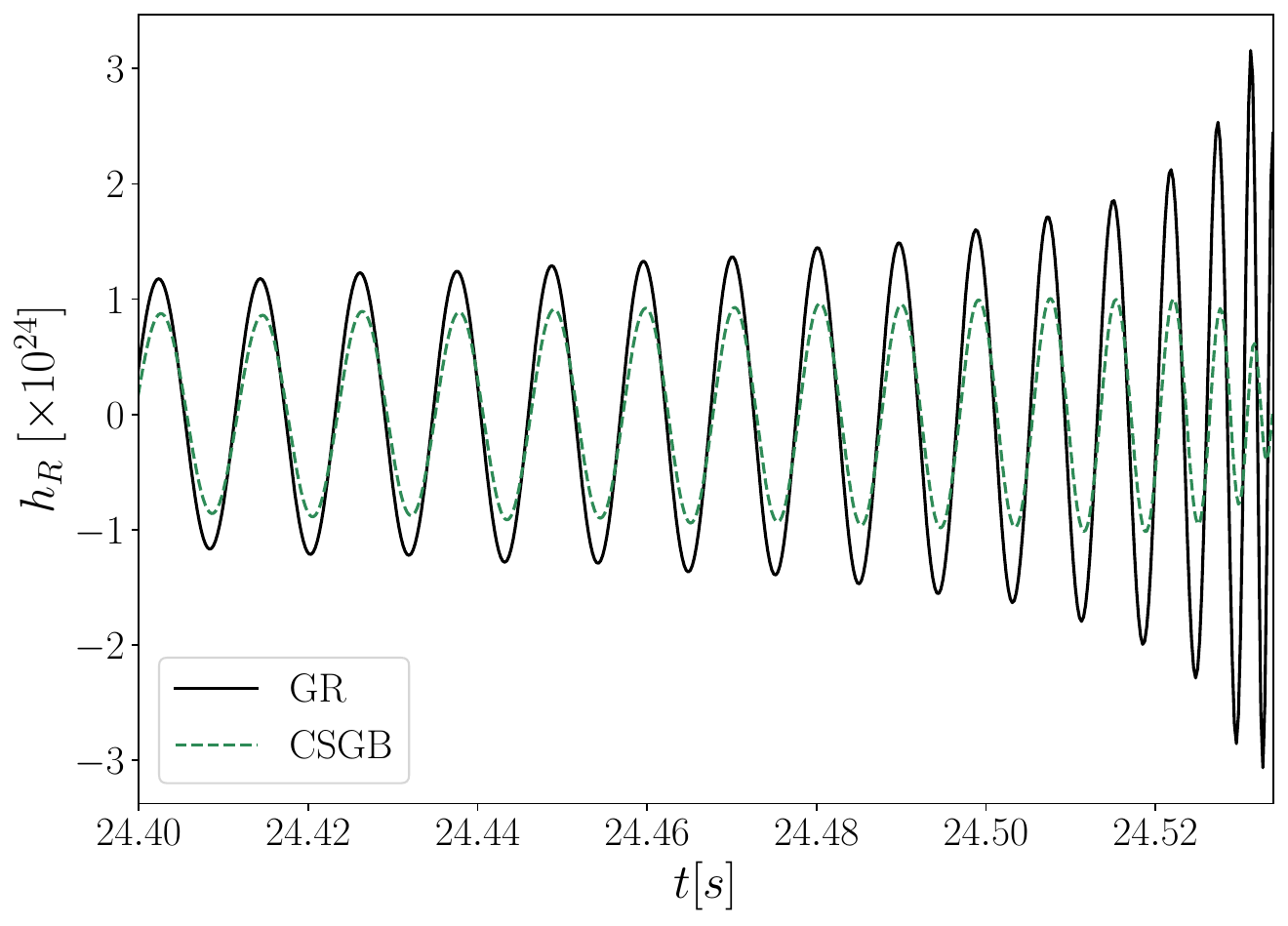}\hfill
    \includegraphics[width=0.5\textwidth]{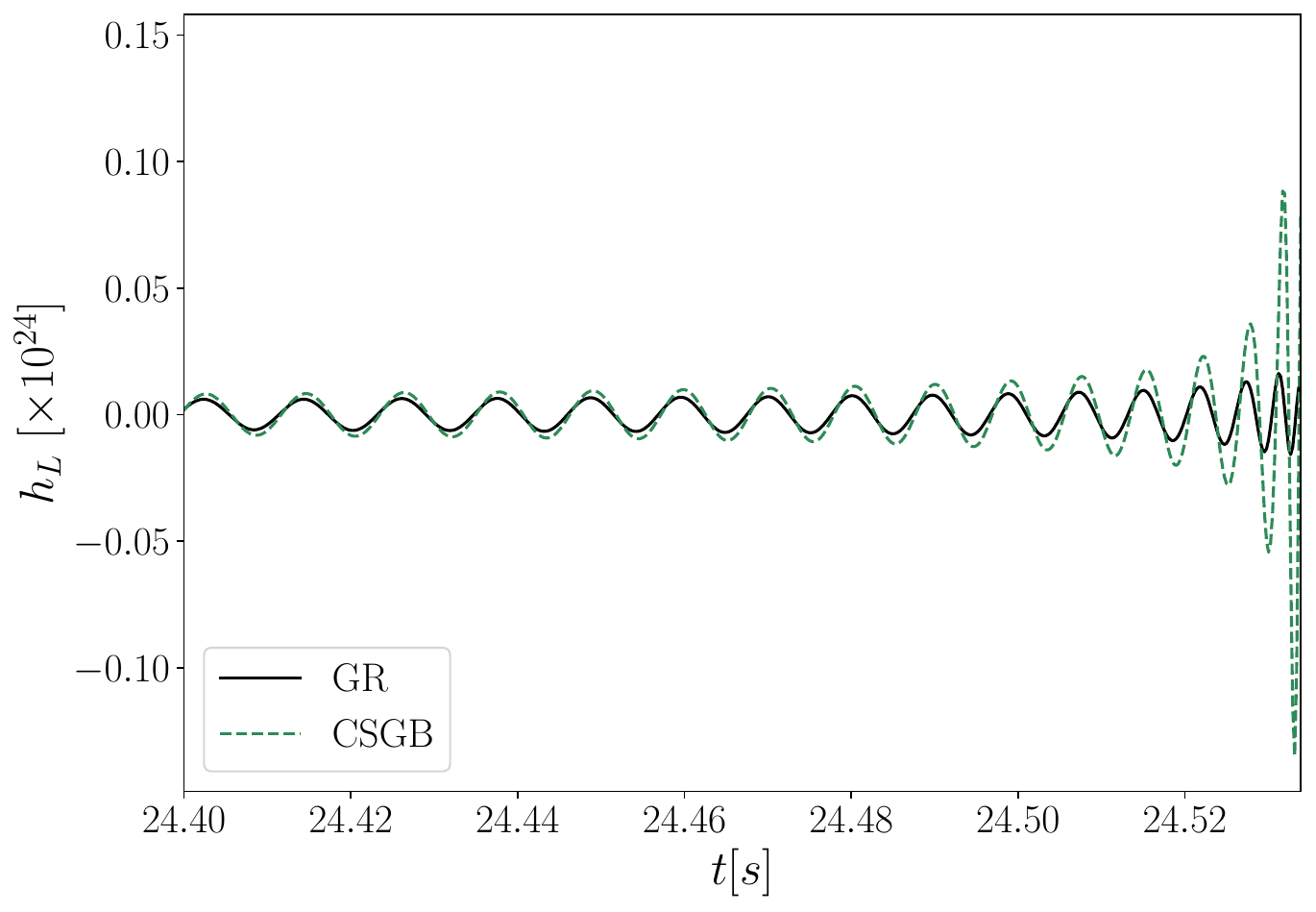}
    \caption{Example modification to a binary black hole waveform for $h_R$ (left) and $h_L$ (right). We see a constant phase shift across both polarizations due to the parity-invariant modification to the phase, and an attenuation/enhancement of the amplitude for $h_R$ and $h_L$, respectively, due to the parity-violating modification to the amplitude. To generate the waveform, we employ the \texttt{GW Analysis Tools} code \cite{Perkins:2021mhb}. For the source parameters we take $m_1 = 20 M_\odot$, $m_2 = 18 M_\odot$, $\iota = 2.6$ rad, $\psi = 3.14$ rad, $RA = 3.45$ rad, $Dec = -3968$ rad. For computational ease we rescale $f/100$ Hz, and $D_2/\text{Gpc}$. The modification parameters are chosen to be artificially large in order to visually see the effects; in dimensionless units $\phi_0' = 3$ and $\varphi_0' = 5$.}
    \label{fig:waveform}
\end{figure*}

We show an example of this modification to the waveform of a binary black hole in Fig.~\ref{fig:waveform}. We can see that both the right and left polarizations have the same phase shift as a result of the parity-invariant correction to the phase. The amplitude attenuates for $h_R$ and is amplified for $h_L$ due to the parity-violating amplitude corrections.

Furthermore, we can see how the GW velocity is modified for CS-GB gravity.  From Eqs.~(\ref{eq:hrl}) and (\ref{eq:c}), the GWs satisfy the dispersion relation
\begin{align}
    \omega_{R,L}^2 = k^2\bigg[1 - \frac{\alpha'}{2}\bigg(\frac{1}{a}\mathcal{H}\phi' + \varphi'^2\bigg)\bigg]. \label{eq:dispersionrelation}
\end{align}
From Eq.~(\ref{eq:dispersionrelation}), we can find the group and phase velocities of a GW, which are given by $v_g = d\omega/dk$ and $v_p = \omega/k$, respectively. We have
\begin{align}
    v_g^{R,L} = v_p^{R,L} = 1 - \frac{\alpha'}{4}\bigg(\frac{1}{a}\mathcal{H}\phi' + \varphi'^2\bigg). \label{eq:vgvp}
\end{align}
In Appendix~\ref{sec:moddispersiongeneral}, we show how Eq.~(\ref{eq:vgvp}) can be generalized for any extension to GR, using the framework that is presented in the next section.

\section{Extension of Parity-Violating Parametrization and Constraints}\label{sec:5}

In this section, we place our work in a broader context by making contact with the parametrization in \cite{Jenks2023} in Sec.~\ref{sec:parametrization} and then discussing observational constraints on the theory in Sec~\ref{sec:constraints}.

\subsection{Parametrization}
\label{sec:parametrization}
 We would like to place our work in the context of the parametrization in \cite{Jenks2023}, in which it was shown that generic parity-violating corrections to the GW propagation equations can be written in a theory-agnostic way using dimensionless parameters; a particular theory will then correspond to specific values of these parameters. A similar parametrization to \cite{Jenks2023} for describing parity-violating propagation effects was also introduced in \cite{Qiao2019, Zhao2020, Zhu2023}.

In our expression Eq.~\eqref{eq:hrl}, because we have contributions from both the CS and GB corrections, we have both parity-even and parity-odd terms. Thus, we can extend the parametrization in \cite{Jenks2023} to also account for parity-invariant terms such that the GW propagation equation can be written as
\begin{widetext}
\begin{align}
    &h_{R,L}''~+~\bigg\{2\mathcal{H}~+~\sum_{n=0}^{\infty}(\lambda_{R,L}k)^n\bigg[\frac{\alpha_n(\eta)}{(\Lambda a)^n}\mathcal{H}~+~\frac{\beta_n(\eta)}{(\Lambda a)^{n-1}}\bigg]\bigg\}h_{R,L}' \nonumber \\ +~&k^2\bigg\{1~+~\sum_{m=0}^{\infty}(\lambda_{R,L})^{m+1}k^{m-1}\bigg[\frac{\gamma_m(\eta)}{(\Lambda a)^m}\mathcal{H}~+~\frac{\delta_m(\eta)}{(\Lambda a)^{m-1}}\bigg]\bigg\}h_{R,L} = 0, \label{eq:generalform-pipv}
\end{align}
\end{widetext}
where $\{\alpha_n, \beta_n, \gamma_m, \delta_m\}$ are the dimensionless parameters that depend on the specific theory in consideration, and $\Lambda$ is the cutoff scale of the theory. When $\alpha = \beta = \gamma = \delta = 0$, we recover the propagation equation for GWs in GR. In CS gravity, for example, the parameter $\alpha_1 \neq 0$ with all other parameters vanishing. 

Here $m$ and $n$ are integers; this extends the parametrization in \cite{Jenks2023} in which $n$ and $m$ were constrained to be odd and even integers, respectively, as to consider only parity-violating effects. With this extension, one can now explicitly see the propagation effects of theories with both parity-violating and parity-invariant contributions. This extension also cleanly maps to ppE \cite{Yunes:2009ke} and can be easily used in data analysis.

Comparing Eq.~(\ref{eq:generalform-pipv}) with Eqs.~(\ref{eq:b}) and (\ref{eq:c}), we can make the identification that $\alpha_1 = -2\alpha'\Tilde{\varphi}'$, $\beta_1 = 2\alpha'\phi'\varphi'$, $\gamma_1 = -\frac{1}{2}\alpha'\Tilde{\phi}'$ and $\delta_1 = -\frac{1}{2}\alpha'\varphi'^2$, with all other parameters being zero, where we have introduced a rescaling of the scalar fields by $\Lambda$ such that $\Tilde{\phi} \equiv \phi\Lambda$ and $\Tilde{\varphi} \equiv \varphi\Lambda$.  

\subsection{Constraints}
\label{sec:constraints}
While a full data analysis will be necessary to rigorously constrain
$\phi'$ and $\varphi'$, as a
first step, we can consider initial constraints based on
previously existing work in the literature. Significant work has been done to constrain birefringent effects from a variety of GW sources, e.g.~\cite{Okounkova2022, Zhao2022, Ng2023, Callister2023, Lagos2024}. Here, we consider both the velocity constraints from
the GW170817/GRB170817 coincident event and birefringence specific constraints in the literature from binary black hole events. 

The coincident GW/gamma ray burst
event from the binary neutron star merger GW170817 has
provided a tight constraint on the speed of GWs, $c_T$, compared to the speed of light, $c$. We have \cite{Abbott2017}
\begin{align}
    -7 \times 10^{-16} < 1 - c_T < 3 \times 10^{-15}. \label{eq:cconstraint}
\end{align}
The constraint in Eq.~(\ref{eq:cconstraint}) rules out many beyond-GR theories that induce a modification to the GW speed \cite{Baker2017, Creminelli2017, Sakstein2017, Ezquiaga2017, Wang2017}. While it has been shown that the GW speed in CS gravity is equal to the speed of light \cite{Nojiri2019}, this is not the case for GB gravity.\footnote{However, with modifications to the scalar GB potential, the GW speed in GB gravity will equal the speed of light \cite{Odintsov2020, Nojiri2024}.} As a result, CS-GB gravity also induces modifications to the GW speed which are thus constrained by Eq.~(\ref{eq:cconstraint}).  We can map this constraint to our parametrization Eq.~(\ref{eq:generalform-pipv}) to constrain the CS-GB theory parameters such that the CS-GB modified GW speed does not violate the observational bound Eq.~(\ref{eq:cconstraint}).

From Eq.~(\ref{eq:vgvp}), we have
\begin{align}
    |v_g - 1| = \frac{\alpha'}{4}\bigg(\frac{1}{a}\mathcal{H}\phi' + \varphi'^2\bigg). \label{eq:135}
\end{align}
Taking the weaker constraint of Eq.~(\ref{eq:cconstraint}), and neglecting the term that is suppressed by $\mathcal{H}/\Lambda_{PV}$ in Eq.~(\ref{eq:135}), we have
\begin{align}
    \bigg|\frac{1}{4}\alpha'\varphi'^2\bigg| < 3 \times 10^{-15}. \label{eq:136}
\end{align}
One can combine Eq.~(\ref{eq:136}) with constraints on $\alpha'$ from GB gravity (see e.g. \cite{Wang2021, Lyu2022}) to obtain a bound on $\varphi'$, which is roughly $\Tilde{\varphi}' \lesssim 10^{-15}~\text{eV}^2$ (in natural units)\footnote{Upon converting Eq.~(\ref{eq:136}) from geometric to natural units ($c = \hbar = 1)$, we multiply $\varphi'$ by the cutoff scale $\Lambda_{PV}$, which has a lower bound $\Lambda_{PV} \gtrsim 10^2$ eV \cite{Jenks2023}. We do the same for $\phi_0'$ to get the constraint on $\Tilde{\phi}_0'$ from Eq.~(\ref{eq:betaconstraint}).}.

We can then use the constraint from \cite{Ng2023} via \cite{Jenks2023},
\begin{align}
    \bigg|2\alpha'\phi_0'\varphi_0'\bigg| < 0.7 \times 10^{-20}, \label{eq:betaconstraint}
\end{align}
and combining Eq.~(\ref{eq:betaconstraint}) with Eq.~(\ref{eq:136}) allows one to place a constraint on $\phi_0'$, which is roughly $\Tilde{\phi}'_0 \lesssim 10^{-22}~\text{eV}^2$.

\section{Discussion and Conclusions}\label{sec:discussion}

In this work, we have studied the propagation of GWs in CS-GB gravity. We have reviewed the derivation of CS-GB gravity from HST and derived how GW propagation is modified in such a theory. We have furthermore extended the parametrization first introduced in \cite{Jenks2023} for the parity-violating sector to include the parity-even sector. The framework presented in this paper thus allows one to study any correction to GR in explicitly parity-violating and parity-invariant contributions. Moreover, we have used this parametrization to map the CS-GB modifications to GW observables, which allows us to place constraints on the theory parameters.

As we have seen, CS-GB gravity (and modified gravity theories in general) will modify both the amplitude and phase of a GW. Most of this paper has focused on these modifications for the propagation of a GW, but these modifications can also arise in the generation of GWs. In compact binary coalescences, the presence of the axion and dilaton will extract energy from the binary, leading to a modification of the chirp mass (see e.g. \cite{Yunes2009, Sopuerta2009, Yagi2012, AlexanderYunes2018}). The two effects can be considered independently, with the generation effects being of $\mathcal{O}(\alpha'^2)$, making them subdominant to the propagation effects, which are of $\mathcal{O}(\alpha')$ \cite{Callister2023}. 

Furthermore, CS-GB gravity can impact GWs during inflation. For example, tensor perturbations of the spacetime metric source primordial GWs, which encode important information of the early Universe and provide an important test of GR. During inflation, the Pontryagin term associated with CS gravity can lead to the resonant amplification of GWs on small scales \cite{Peng2022}, and one can study the energy spectrum associated with these GWs \cite{Odintsov2022}. It would be interesting to determine how these scenarios would be modified in CS-GB gravity. We leave this study for future work.\footnote{Primordial GWs arising from CS-GB gravity have been previously studied in \cite{Satoh2008}, but with a single scalar field associated to both the CS and GB terms instead of two separate scalar fields, like we are considering in this paper.}

The gravitational field in the exterior of supermassive, spinning black holes (BHs) is crucial in the emission of GWs. In GR, such a field is described by the Kerr metric \cite{Kerr1963}, which is a stationary and axisymmetric solution, parametrized in terms of the mass of the BH and its angular momentum. However, in modified theories of gravity, the Kerr metric does not need to be a solution to the field equations. For example, in the case of GB gravity, slowly rotating BH solutions have been found that differ from Kerr \cite{Pani2009}. A measured deviation from the Kerr metric, whether from electromagnetic or GW observations \cite{Psaltis2007, Psaltis2008, Glampedakis2006, Collins2004}, can therefore provide insight into extensions of GR, or lack thereof. 

One can ask what metric represents a spinning BH in CS-GB gravity. For CS, the metric and scalar field perturbations describing the leading order corrections to the Kerr metric are known \cite{Yunes2009, Delsate2018, Chatzifotis2022}. The leading order corrections for GB have been analyzed as well \cite{Kleihaus2011, Maselli2015, Kleihaus2016}. We leave an in-depth analysis of the BH solution in CS-GB gravity for future work.

\acknowledgments
The authors thank Heliudson Bernardo and Tucker Manton for helpful comments and discussion, including reviewing an early draft. T.D. and S.A. are supported by the Simons Foundation, Award 896696. L.~J.~ is supported by the Kavli Institute for Cosmological Physics at the University of Chicago via an endowment from the Kavli Foundation and its founder Fred Kavli. 

\appendix
\section{Stringy Derivation} \label{sec:stringyderivation}
In this appendix, we provide more details on the derivation of the 4D effective action Eq.~(\ref{eq:action}) in Sec.~\ref{sec:2}, following \cite{Cano2021}. 

With the CS term in the 10D heterotic superstring effective action (Eq.~(\ref{eq:heterotic})), $\hat{H}$ satisfies the modified Bianchi identity, 
\begin{align}
    d\hat{H} = \frac{\alpha'}{4}\hat{R}_{(-)~~b}^{~~~~a} \wedge \hat{R}_{(-)~a}^{~~b}. \label{eq:bianchi}
\end{align}
Upon compactifying Eq.~(\ref{eq:heterotic}) on a six-torus in Eq.~(\ref{eq:compactification}), Eq.~(\ref{eq:bianchi}) can be written as
\begin{align}
    \frac{1}{3!}\epsilon^{\mu\nu\rho\sigma}\Bar{\nabla}_{\mu}H_{\nu\rho\sigma} + \frac{\alpha'}{8}\Bar{R}_{(-)\nu\rho\sigma}\Tilde{\Bar{R}}_{(-)}^{\mu\nu\rho\sigma} = 0,
\end{align}
where 
\begin{align}
    \Tilde{\Bar{R}}_{(-)}^{\mu\nu\rho\sigma} = \frac{1}{2}\epsilon^{\mu\nu\alpha\beta}\Bar{R}_{(-)\alpha\beta}^{~~~~~~\rho\sigma}.
\end{align}
After integrating by parts, we get Eq.~(\ref{eq:4dcompact}):
\begin{align}
    \Bar{S} &= \frac{1}{16\pi G_N^{(4)}}\int d^4x\sqrt{|\Bar{g}|}\bigg\{e^{-2(\hat{\phi} - \hat{\phi}_{\infty})}\bigg[\Bar{R} - 4(\partial\hat\phi)^2 \nonumber \\ &+ \frac{1}{12}H^2\bigg] - \frac{1}{3!}H_{\mu\nu\rho}\epsilon^{\mu\nu\rho\sigma}\partial_{\sigma}\varphi + \frac{\alpha'}{8}\mathcal{L}_{R^2} + \mathcal{O}(\alpha'^3)\bigg\}, \label{eq:4dcompact2}
\end{align}
where 
\begin{align}
    \mathcal{L}_{R^2} = e^{-2(\hat{\phi} - \hat{\phi}_{\infty})}\Bar{R}_{(-)\mu\nu\rho\sigma}\Bar{R}_{(-)}^{\mu\nu\rho\sigma} - \varphi\Bar{R}_{(-)\mu\nu\rho\sigma}\Tilde{\Bar{R}}_{(-)}^{{\mu\nu\rho\sigma}}.
\end{align}
Now, from the variation of $H$, we have that
\begin{align}
    e^{-2(\hat{\phi} - \hat{\phi}_{\infty})}\frac{1}{6}H_{\mu\nu\rho} - \frac{1}{6}\epsilon_{\mu\nu\rho\sigma}\Bar{\nabla}^{\sigma}\varphi + \frac{\alpha'}{8}\frac{\delta\mathcal{L}_{R^2}}{\delta H^{\mu\nu\rho}} = 0,
\end{align}
and as explained in Section \ref{sec:2}, to solve it we expand $H$ in $\alpha'$:
\begin{align}
    H = H^{(0)} + \alpha'H^{(1)} + \alpha'^2H^{(2)} + ..., \label{eq:hexp}
\end{align}
which after plugging the expansion Eq.~(\ref{eq:hexp}) into Eq.~(\ref{eq:4dcompact2}), we arrive at Eq.~(\ref{eq:jordanaction}), 
\begin{align}
    \Bar{S} &= \frac{1}{16\pi G_N^{(4)}}\int d^4x\bigg\{e^{-2(\hat{\phi}-\hat{\phi}_{\infty})}\bigg[\Bar{R} - 4(\partial\hat{\phi})^2\bigg] \nonumber \\ &+ \frac{1}{2}e^{2(\hat{\phi}-\hat{\phi}_{\infty})}(\partial\varphi)^2 + \frac{\alpha'}{8}\mathcal{L}_{R^2}\biggr\rvert_{H^{(0)}} + \mathcal{O}(\alpha'^2)\bigg\}\label{eq:jordanaction2}.
\end{align}
To evaluate the four-derivative term $\mathcal{L}_{R^2}$, we have to substitute in the expression for $H^{(0)}$, which is
\begin{align}
    H_{\mu\nu\rho}^{(0)}  = e^{2(\hat{\phi} - \hat{\phi}_{\infty})}\epsilon_{\mu\nu\rho\sigma}\nabla^{\sigma}\varphi,
\end{align}
and use the fact that the curvature $\hat{R}_{(-)}$ can be written in terms of $\hat{H}$ as well as the Riemannian curvature $\hat{R}$:
\begin{align}
    \hat{R}_{(-)\mu\nu~\sigma}^{~~~~~~\rho} &= \hat{R}_{\mu\nu~\sigma}^{~~~\rho} - \hat{\nabla}_{[\mu}\hat{H}_{\nu]~~\sigma}^{~~\rho} - \frac{1}{2}\hat{H}_{[\mu|~\alpha}^{~\rho}\hat{H}_{|\nu]~\sigma}^{~~\alpha}.
\end{align}
Evaluation of the four-derivative term yields Eq.~(\ref{eq:4derivativeterm}),
\begin{align}
    \mathcal{L}_{R^2}\biggr\rvert_{H^{(0)}} &= e^{-2(\hat{\phi}-\hat{\phi}_{\infty})}\bigg[\Bar{R}_{\mu\nu\rho\sigma}\Bar{R}^{\mu\nu\rho\sigma} + 6\Bar{G}_{\mu\nu}A^{\mu}A^{\nu} + \frac{7}{4}A^4 \nonumber \\ &- 2\Bar{\nabla}_{\mu}A_{\nu}\Bar{\nabla}^{\mu}A^{\nu} - (\Bar{\nabla}_{\mu}A^{\mu})^2\bigg] - \varphi\Bar{R}_{\mu\nu\rho\sigma}\Tilde{\Bar{R}}^{\mu\nu\rho\sigma} \nonumber \\ &+ \text{total derivatives}. \label{eq:4derivativeterm2}
\end{align}
Upon transforming our theory from the Jordan frame to the Einstein frame via the conformal rescaling Eq.~(\ref{eq:conformalrescaling}), the effect on the two-derivative terms in the Lagrangian is rather straightforward to compute:
\begin{align}
    \sqrt{|\Bar{g}|}\mathcal{L}_2 = \sqrt{|g|}\bigg[R + 2(\partial\hat{\phi})^2 + \frac{1}{2}e^{4(\hat{\phi} - \hat{\phi}_{\infty})}(\partial\varphi)^2\bigg].
\end{align}
On the other hand, the effect of the conformal rescaling on the four-derivative term $\mathcal{L}_{R^2}$ requires a lengthier calculation; we need to take into account the transformation of the Riemann tensor and the covariant derivative, and integrate by parts multiple times. The end result is
\begin{align}
    \sqrt{|\Bar{g}|}&\mathcal{L}_{R^2}\biggr\rvert_{H^{(0)}} = \sqrt{|g|}\bigg\{e^{-2(\hat{\phi} - \hat{\phi}_{\infty}})\bigg[R_{\mu\nu\rho\sigma}R^{\mu\nu\rho\sigma} \nonumber \\ +~&4R^{\mu\nu}(4\partial_{\mu}\hat{\phi}\partial_{\nu}\hat{\phi} + A_{\mu}A_{\nu}) + R[4\nabla^2\hat{\phi} - 4(\partial\hat{\phi})^2 - 3A^2] \nonumber \\ +~&12(\partial\hat{\phi})^4 + 12(\nabla^2\hat{\phi})^2 + \frac{7}{4}A^4 - 12(\partial_{\mu}\hat{\phi}A^{\mu})^2 \nonumber \\ -~&2A^2(\partial\hat{\phi})^2 - 8A^2\nabla^2\hat{\phi} - 16\partial_{\mu}\hat{\phi}A^{\mu}\nabla_{\alpha}A^{\alpha} - 3(\nabla_{\alpha}A^{\alpha})^2\bigg] \nonumber \\ -~&\varphi R_{\mu\nu\rho\sigma}\Tilde{R}^{\mu\nu\rho\sigma}\bigg\} + \text{total derivatives},
\end{align}
which we can rewrite as
\begin{align}
    \sqrt{\overline{\Bar{g}}}\mathcal{L}_{R^2}\biggr\rvert_{H^{(0)}} = \sqrt{\Bar{g}}\bigg[e^{-2(\hat{\phi} - \hat{\phi}_{\infty})}\mathcal{X}_4 - \varphi R_{\mu\nu\rho\sigma}\Tilde{R}^{\mu\nu\rho\sigma} + \mathcal{L}'\bigg],
\end{align}
where $\mathcal{X}_4 = R^2 - 4R_{\mu\nu}R^{\mu\nu} + R_{\mu\nu\rho\sigma}R^{\mu\nu\rho\sigma}$ is the 4D GB density, and we have collected the remaining terms in $\mathcal{L}'$. 

Now, let us consider the zeroth order equations of motion
\begin{align}
    \mathcal{E}_{\mu\nu} &= R_{\mu\nu} + 2\partial_{\mu}\hat{\phi}\partial_{\nu}\hat{\phi} + \frac{1}{2}A_{\mu}A_{\nu}, \label{eq:emunueom}\\
    \mathcal{E}_{\hat{\phi}} &= \nabla^2\hat{\phi} - \frac{1}{2}A^2, \label{eq:ephieom} \\
    \mathcal{E}_{\varphi} &= \nabla_{\mu}A^{\mu} + 2\partial_{\mu}\hat{\phi}A^{\mu}. \label{eq:evarphieom}
\end{align}
After some algebra, $\mathcal{L}'$ can be written in terms of Eqs.~(\ref{eq:emunueom})-(\ref{eq:evarphieom}) as follows:
\begin{align}
    \mathcal{L}' &= e^{-2(\hat{\phi} - \hat{\phi}_{\infty})}\bigg\{4\mathcal{E}_{\mu\nu}\mathcal{E}^{\mu\nu} - \mathcal{E}^2 + 12\mathcal{E}_{\hat{\phi}}^2 + 4\mathcal{E}\mathcal{E}_{\hat{\phi}} - 3\mathcal{E}_{\varphi}^2 \nonumber \\ &+ 2\mathcal{E}_{\hat{\phi}}[A^2 - 4(\partial\hat{\phi})^2] - 4\mathcal{E}_{\varphi}\partial_{\mu}\hat{\phi}A^{\mu}\bigg\}. \label{eq:l'}
\end{align}
We see that all the terms in $\mathcal{L}'$ are proportional to the zeroth-order equations of motion, which means if we redefine the fields
\begin{align}
    g_{\mu\nu} &\rightarrow g_{\mu\nu} + \alpha'\Delta_{\mu\nu}, \\ \hat{\phi} &\rightarrow \hat{\phi} + \alpha'\Delta\hat{\phi}, \\ \varphi &\rightarrow \varphi + \alpha'\Delta\varphi,
\end{align}
then we introduce terms linear in $\alpha'$ that are proportional to the zeroth order equations of motion, which we can therefore use to cancel all the terms in $\mathcal{L}'$ \cite{Cano2021}.

Thus, introducing the 4D dilaton $\phi = 2(\hat{\phi} - \hat{\phi}_{\infty})$, we end up with Eq.~(\ref{eq:action}), a very simple form of our action in four dimensions:
\begin{align}
    S &= \frac{1}{16\pi}\int d^4x\sqrt{|g|}\bigg[R + \frac{1}{2}(\partial\phi)^2 + \frac{1}{2}e^{2\phi}(\partial\varphi)^2 \nonumber \\ &+ \frac{\alpha'}{8}\bigg(e^{-\phi}\mathcal{X}_4 - \varphi R_{\mu\nu\rho\sigma}\Tilde{R}^{\mu\nu\rho\sigma}\bigg) + \mathcal{O}(\alpha'^2)\bigg]. \label{eq:action2}
\end{align}

\section{Taylor Expansion for GW Propagation Coefficients} \label{sec:gwcoefftaylor}
Here we show the steps in expanding the $\Bar{B}$ and $\Bar{C}$ coefficients in Eq.~(\ref{eq:hrl}) to linear order in $\alpha'$. 

For $\Bar{B}$, we have
\begin{align}
    \Bar{B} &\approx \bigg(1 + \frac{\alpha'}{4a^2}\phi'' + \frac{\alpha'a^2}{2}\lambda_{R,L}k\varphi'\bigg)\bigg[2\mathcal{H} + \frac{\alpha'}{2a^2}\mathcal{H}\phi'' \nonumber \\ &- \lambda_{R,L}\alpha'k\bigg(a^2\mathcal{H}\varphi' + \frac{\varphi''}{a^2}\bigg)\bigg] \\
    &\approx 2\mathcal{H} + \frac{\alpha'}{a^2}\mathcal{H}\phi'' - \frac{\alpha'}{a^2}k\lambda_{R,L}\varphi'',
\end{align}
where we have assumed that $k \gg \mathcal{H}$. We can then use Eqs.~(\ref{eq:phibkgd}) and (\ref{eq:varphibkgd}) to obtain
\begin{align}
     \Bar{B} &= 2\mathcal{H} - \lambda_{R,L}k\frac{2\alpha'}{a^2}\bigg(\mathcal{H}\varphi' - \phi'\varphi'\bigg). \label{eq:b3}
\end{align}
Now, we need to correct for the factors of $a$, since $(1/a)(da/d\eta) = da/dt$. Thus, the conformal time derivatives in Eq.~(\ref{eq:b3}) pick up an extra factor of $a$. So, we have
\begin{align}
    \Bar{B} &= 2\mathcal{H} - 2\lambda_{R,L}\alpha'k\bigg(\frac{\mathcal{H}}{a}\varphi' - \phi'\varphi'\bigg),
\end{align}
which is Eq.~(\ref{eq:b}).

For $\Bar{C}$, we have
\begin{align}
    \Bar{C} &\approx \bigg(1 + \frac{\alpha'}{4a^2}\phi'' + \lambda_{R,L}\frac{\alpha'a^2}{2}k\varphi'\bigg)\bigg\{k^2 - 2\mathcal{H}^2 + 6\mathcal{H}' \nonumber \\ &+ \frac{\alpha'}{2a^2}\phi''\bigg(4\mathcal{H}^2 - 12\mathcal{H}' - k^2\bigg) \nonumber \\ &- \lambda_{R,L}\frac{\alpha'}{2}k\bigg[a^2\varphi'(4\mathcal{H}^2 + k^2) + \frac{2\mathcal{H}\varphi''}{a^4}\bigg]\bigg\} \label{eq:cexp1} \\
    &\approx k^2 - \frac{\alpha'k^2}{4a^2}\phi'', \label{eq:cexp2}
\end{align}
where in going from Eq.~(\ref{eq:cexp1}) to (\ref{eq:cexp2}) we have again assumed that $k \gg \mathcal{H}$. Furthermore, we can assume that $\phi$ and $\varphi$ are small to retain terms that are at most second-order in the scalar fields in Eq.~(\ref{eq:cexp2}).

Plugging Eq.~(\ref{eq:phibkgd}) into Eq.~(\ref{eq:cexp2}) yields
\begin{align}
    \Bar{C} = k^2\bigg[1 - \frac{\alpha'}{2a^2}\bigg(\mathcal{H}\phi' + \varphi'^2\bigg)\bigg],
\end{align}
and again noting that $(1/a)(da/d\eta) = da/dt$ to correct the factors of $a$ in the conformal time derivatives, we end up with 
\begin{align}
    \Bar{C} &= k^2\bigg[1 - \frac{\alpha'}{2}\bigg(\frac{\mathcal{H}}{a}\phi' + \varphi'^2\bigg)\bigg],
\end{align}
which is Eq.~(\ref{eq:c}).

\section{Generalization of Modified Dispersion Relation}\label{sec:moddispersiongeneral}
The discussion in Sec.~\ref{sec:4} from Eq.~(\ref{eq:moddispersion}) onwards can be generalized for any modification to GR by extending the discussion in \cite{Jenks2023} to include the parity-even sector. From Eq.~(19) of \cite{Jenks2023} and Eq.~(\ref{eq:generalform-pipv}), it is straightforward to see that the effective modified dispersion relation Eq.~(\ref{eq:moddispersion}) can be parametrized as
\begin{align}
    \theta'' &+ \theta'^2 + i\theta'\bigg\{2\mathcal{H} + (\lambda_{R,L}k)^n\bigg[\frac{\alpha_n}{(\Lambda_{PV}a)^n}\mathcal{H} + \frac{\beta_n}{(\Lambda_{PV}a)^{n-1}}\bigg]\bigg\} \nonumber \\ &- k^2\bigg\{1 + (\lambda_{R,L})^{m+1}k^{m-1}\bigg[\frac{\gamma_m}{(\Lambda_{PV}a)^m}\mathcal{H} + \frac{\delta_m}{(\Lambda_{PV}a)^{m-1}}\bigg]\bigg\} \nonumber \\ &= 0, 
\end{align}
where we are keeping the sums over $n$ and $m$ implicit.

From Eq.~(20) of \cite{Jenks2023}, we can see that the generalization of Eq.~(\ref{eq:birefringence}) is
\begin{align}
    \delta\theta = -i(\lambda_{R,L})^n\delta\theta_A + (\lambda_{R,L})^{m+1}\delta\theta_V,
\end{align}
where the amplitude and velocity birefringence contributions are
\begin{align}
    \delta\theta_A' &= \frac{k^n}{2}\bigg[\frac{\alpha_n}{(\Lambda_{PV}a)^n}\mathcal{H} + \frac{\beta_n}{(\Lambda_{PV}a)^{n-1}}\bigg], \label{eq:general-ampbiref}\\
    \delta\theta_V' &= \frac{k^m}{2}\bigg[\frac{\gamma_m}{(\Lambda_{PV}a)^m}\mathcal{H} + \frac{\delta_m}{(\Lambda_{PV}a)^{m-1}}\bigg]. \label{eq:general-velbiref}
\end{align}
Eqs.~(\ref{eq:general-ampbiref}) and (\ref{eq:general-velbiref}) can be rewritten as
\begin{align}
    \delta\theta_A &= \frac{[k(1+z)]^n}{2}\bigg(\frac{\alpha_{n_0}}{\Lambda_{PV}^n}z_n + \frac{\beta_{n_0}}{\Lambda_{PV}^{n-1}}D_{n+1}\bigg), \\
    \delta\theta_V &= \frac{[k(1+z)^m]}{2}\bigg(\frac{\gamma_{m_0}}{\Lambda_{PV}^m}z_m + \frac{\delta_{m_0}}{\Lambda_{PV}^{m-1}}D_{m+1}\bigg),
\end{align}
such that the right and left-handed polarization modes are modified in the following way
\begin{align}
    &h_{R,L} = \Bar{h}_{R,L} \nonumber \\ &\times \text{exp}\bigg\{-(\lambda_{R,L})^n\frac{[k(1+z)]^n}{2}\bigg(\frac{\alpha_{n_0}}{\Lambda_{PV}^n}z_n + \frac{\beta_{n_0}}{\Lambda_{PV}^{n-1}}D_{n+1}\bigg)\bigg\} \nonumber \\ &\times \text{exp}\bigg\{i(\lambda_{R,L})^{m+1}\frac{[k(1+z)]^m}{2}\bigg(\frac{\gamma_{m_0}}{\Lambda_{PV}^m}z_m + \frac{\delta_{m_0}}{\Lambda_{PV}^{m-1}}D_{m+1}\bigg)\bigg\},
\end{align}
where $\Bar{h}_{R,L}$ is the usual GR expression for the right and left-handed modes.

Via the generalized modified dispersion relation
\begin{align}
    \omega_{R,L}^2 = k^2\bigg\{1 + (\lambda_{R,L})^{m+1}k^{m-1}\bigg[\frac{\gamma_m}{(\Lambda_{PV}a)^m}\mathcal{H} + \frac{\delta_m}{(\Lambda_{PV}a)^{m-1}}\bigg]\bigg\},
\end{align}
the modified group and phase velocities are then
\begin{align}
    v_g^{R,L} &= 1 + \frac{(\lambda_{R,L})^{m+1}}{2}mk^{m-1}\bigg[\frac{\gamma_m}{(a\Lambda_{PV})^m}\mathcal{H} + \frac{\delta_m}{(a\Lambda_{PV})^{m-1}}\bigg], \\
    v_p^{R,L} &= 1 + \frac{(\lambda_{R,L})^{m+1}}{2}k^{m-1}\bigg[\frac{\gamma_m}{(a\Lambda_{PV})^m}\mathcal{H} + \frac{\delta_m}{(a\Lambda_{PV})^{m-1}}\bigg].
\end{align}

\bibliography{bibliography}

@preamble{"\providecommand{\noopsort}[1]{}" #
   "\providecommand{\singleletter}[1]{#1}%"}

@article{Jenks2023,
  title = {Parametrized parity violation in gravitational wave propagation},
  author = {Jenks, Leah and Choi, Lyla and Lagos, Macarena and Yunes, Nicol\'as},
  journal = {Phys. Rev. D},
  volume = {108},
  issue = {4},
  pages = {044023},
  numpages = {19},
  year = {2023},
  month = {Aug},
  publisher = {American Physical Society},
  doi = {10.1103/PhysRevD.108.044023},
  url = {https://link.aps.org/doi/10.1103/PhysRevD.108.044023}
}

@article{Cano2021,
  title = {String gravity in $D=4$},
  author = {Cano, Pablo A. and Ruip\'erez, Alejandro},
  journal = {Phys. Rev. D},
  volume = {105},
  issue = {4},
  pages = {044022},
  numpages = {15},
  year = {2022},
  month = {Feb},
  publisher = {American Physical Society},
  doi = {10.1103/PhysRevD.105.044022},
  url = {https://link.aps.org/doi/10.1103/PhysRevD.105.044022}
}

@article{Alexander2009,
   title={Chern–Simons modified general relativity},
   volume={480},
   ISSN={0370-1573},
   url={http://dx.doi.org/10.1016/j.physrep.2009.07.002},
   DOI={10.1016/j.physrep.2009.07.002},
   number={1–2},
   journal={Physics Reports},
   publisher={Elsevier BV},
   author={Alexander, Stephon and Yunes, Nicolás},
   year={2009},
   month=aug, pages={1–55}}

@article{Odintsov2019,
   title={Inflationary phenomenology of Einstein Gauss-Bonnet gravity compatible with GW170817},
   volume={797},
   ISSN={0370-2693},
   url={http://dx.doi.org/10.1016/j.physletb.2019.134874},
   DOI={10.1016/j.physletb.2019.134874},
   journal={Physics Letters B},
   publisher={Elsevier BV},
   author={Odintsov, S.D. and Oikonomou, V.K.},
   year={2019},
   month=oct, pages={134874} }

@article{Lue1999,
   title={Cosmological Signature of New Parity-Violating Interactions},
   volume={83},
   ISSN={1079-7114},
   url={http://dx.doi.org/10.1103/PhysRevLett.83.1506},
   DOI={10.1103/physrevlett.83.1506},
   number={8},
   journal={Physical Review Letters},
   publisher={American Physical Society (APS)},
   author={Lue, Arthur and Wang, Limin and Kamionkowski, Marc},
   year={1999},
   month=aug, pages={1506–1509} }

@article{Jackiw2003,
  title = {Chern-Simons modification of general relativity},
  author = {Jackiw, R. and Pi, S.-Y.},
  journal = {Phys. Rev. D},
  volume = {68},
  issue = {10},
  pages = {104012},
  numpages = {10},
  year = {2003},
  month = {Nov},
  publisher = {American Physical Society},
  doi = {10.1103/PhysRevD.68.104012},
  url = {https://link.aps.org/doi/10.1103/PhysRevD.68.104012}
}

@article{Conroy2019,
   title={Parity-violating gravity and GW170817 in non-Riemannian cosmology},
   volume={2019},
   ISSN={1475-7516},
   url={http://dx.doi.org/10.1088/1475-7516/2019/12/016},
   DOI={10.1088/1475-7516/2019/12/016},
   number={12},
   journal={Journal of Cosmology and Astroparticle Physics},
   publisher={IOP Publishing},
   author={Conroy, Aindriú and Koivisto, Tomi},
   year={2019},
   month=dec, pages={016–016} }

@article{Horava2009,
  title = {Quantum gravity at a Lifshitz point},
  author = {Ho\ifmmode \check{r}\else \v{r}\fi{}ava, Petr},
  journal = {Phys. Rev. D},
  volume = {79},
  issue = {8},
  pages = {084008},
  numpages = {15},
  year = {2009},
  month = {Apr},
  publisher = {American Physical Society},
  doi = {10.1103/PhysRevD.79.084008},
  url = {https://link.aps.org/doi/10.1103/PhysRevD.79.084008}
}

@article{Zhu2013,
  title = {Effects of parity violation on non-Gaussianity of primordial gravitational waves in Ho\ifmmode \check{r}\else \v{r}\fi{}ava-Lifshitz gravity},
  author = {Zhu, Tao and Zhao, Wen and Huang, Yongqing and Wang, Anzhong and Wu, Qiang},
  journal = {Phys. Rev. D},
  volume = {88},
  issue = {6},
  pages = {063508},
  numpages = {9},
  year = {2013},
  month = {Sep},
  publisher = {American Physical Society},
  doi = {10.1103/PhysRevD.88.063508},
  url = {https://link.aps.org/doi/10.1103/PhysRevD.88.063508}
}

@article{Alexander2006,
  title = {Leptogenesis from Gravity Waves in Models of Inflation},
  author = {Alexander, Stephon H. S. and Peskin, Michael E. and Sheikh-Jabbari, M. M.},
  journal = {Phys. Rev. Lett.},
  volume = {96},
  issue = {8},
  pages = {081301},
  numpages = {4},
  year = {2006},
  month = {Feb},
  publisher = {American Physical Society},
  doi = {10.1103/PhysRevLett.96.081301},
  url = {https://link.aps.org/doi/10.1103/PhysRevLett.96.081301}
}

@article{AlexanderGates2006,
   title={Can the string scale be related to the cosmic baryon asymmetry?},
   volume={2006},
   ISSN={1475-7516},
   url={http://dx.doi.org/10.1088/1475-7516/2006/06/018},
   DOI={10.1088/1475-7516/2006/06/018},
   number={06},
   journal={Journal of Cosmology and Astroparticle Physics},
   publisher={IOP Publishing},
   author={Alexander, Stephon H S and Gates, S James},
   year={2006},
   month=jun, pages={018–018} }

@book{Polchinski1998,
    author = "Polchinski, J.",
    title = "{String theory. Vol. 2: Superstring theory and beyond}",
    doi = "10.1017/CBO9780511618123",
    isbn = "978-0-511-25228-0, 978-0-521-63304-8, 978-0-521-67228-3",
    publisher = "Cambridge University Press",
    series = "Cambridge Monographs on Mathematical Physics",
    month = "12",
    year = "2007"
}

@article{Zwiebach1985,
    author = "Zwiebach, Barton",
    title = "{Curvature Squared Terms and String Theories}",
    reportNumber = "UCB-PTH-85/10",
    doi = "10.1016/0370-2693(85)91616-8",
    journal = "Phys. Lett. B",
    volume = "156",
    pages = "315--317",
    year = "1985"
}

@article{Deser1986,
    author = "Deser, Stanley and Redlich, A. N.",
    title = "{String Induced Gravity and Ghost Freedom}",
    reportNumber = "BRX-TH-200",
    doi = "10.1016/0370-2693(86)90177-2",
    journal = "Phys. Lett. B",
    volume = "176",
    pages = "350",
    year = "1986",
    note = "[Erratum: Phys.Lett.B 186, 461 (1987)]"
}

@article{Yunes2009,
  title = {Dynamical Chern-Simons modified gravity: Spinning black holes in the slow-rotation approximation},
  author = {Yunes, Nicol\'as and Pretorius, Frans},
  journal = {Phys. Rev. D},
  volume = {79},
  issue = {8},
  pages = {084043},
  numpages = {14},
  year = {2009},
  month = {Apr},
  publisher = {American Physical Society},
  doi = {10.1103/PhysRevD.79.084043},
  url = {https://link.aps.org/doi/10.1103/PhysRevD.79.084043}
}

@article{Boulware1986,
    author = "Boulware, David G. and Deser, Stanley",
    title = "{Effective Gravity Theories With Dilatons}",
    reportNumber = "NSF-ITP-86-57",
    doi = "10.1016/0370-2693(86)90614-3",
    journal = "Phys. Lett. B",
    volume = "175",
    pages = "409--412",
    year = "1986"
}

@article{Kanti1996,
  title = {Dilatonic black holes in higher curvature string gravity},
  author = {Kanti, P. and Mavromatos, N. E. and Rizos, J. and Tamvakis, K. and Winstanley, E.},
  journal = {Phys. Rev. D},
  volume = {54},
  issue = {8},
  pages = {5049--5058},
  numpages = {0},
  year = {1996},
  month = {Oct},
  publisher = {American Physical Society},
  doi = {10.1103/PhysRevD.54.5049},
  url = {https://link.aps.org/doi/10.1103/PhysRevD.54.5049}
}

@article{Torii1997,
  title = {Dilatonic black holes with a Gauss-Bonnet term},
  author = {Torii, Takashi and Yajima, Hiroki and Maeda, Kei-ichi},
  journal = {Phys. Rev. D},
  volume = {55},
  issue = {2},
  pages = {739--753},
  numpages = {0},
  year = {1997},
  month = {Jan},
  publisher = {American Physical Society},
  doi = {10.1103/PhysRevD.55.739},
  url = {https://link.aps.org/doi/10.1103/PhysRevD.55.739}
}

@article{Alexeyev1997,
  title = {Black hole solutions with dilatonic hair in higher curvature gravity},
  author = {Alexeyev, S. O. and Pomazanov, M. V.},
  journal = {Phys. Rev. D},
  volume = {55},
  issue = {4},
  pages = {2110--2118},
  numpages = {0},
  year = {1997},
  month = {Feb},
  publisher = {American Physical Society},
  doi = {10.1103/PhysRevD.55.2110},
  url = {https://link.aps.org/doi/10.1103/PhysRevD.55.2110}
}

@article{Moura2006,
   title={Higher-derivative-corrected black holes: perturbative stability and absorption cross section in heterotic string theory},
   volume={24},
   ISSN={1361-6382},
   url={http://dx.doi.org/10.1088/0264-9381/24/2/006},
   DOI={10.1088/0264-9381/24/2/006},
   number={2},
   journal={Classical and Quantum Gravity},
   publisher={IOP Publishing},
   author={Moura, Filipe and Schiappa, Ricardo},
   year={2006},
   month=dec, pages={361–386} }

@article{Guo2008,
   title={Black Holes in the Dilatonic Einstein-Gauss-Bonnet Theory in Various Dimensions. I: -- Asymptotically Flat Black Holes --},
   volume={120},
   ISSN={1347-4081},
   url={http://dx.doi.org/10.1143/PTP.120.581},
   DOI={10.1143/ptp.120.581},
   number={3},
   journal={Progress of Theoretical Physics},
   publisher={Oxford University Press (OUP)},
   author={Guo, Z.-K. and Ohta, N. and Torii, T.},
   year={2008},
   month=sep, pages={581–607} }

@article{Maeda2009,
   title={Black hole solutions in string theory with Gauss-Bonnet curvature correction},
   volume={80},
   ISSN={1550-2368},
   url={http://dx.doi.org/10.1103/PhysRevD.80.104032},
   DOI={10.1103/physrevd.80.104032},
   number={10},
   journal={Physical Review D},
   publisher={American Physical Society (APS)},
   author={Maeda, Kei-ichi and Ohta, Nobuyoshi and Sasagawa, Yukinori},
   year={2009},
   month=nov }

@article{Pani2009,
   title={Are black holes in alternative theories serious astrophysical candidates? The case for Einstein-dilaton-Gauss-Bonnet black holes},
   volume={79},
   ISSN={1550-2368},
   url={http://dx.doi.org/10.1103/PhysRevD.79.084031},
   DOI={10.1103/physrevd.79.084031},
   number={8},
   journal={Physical Review D},
   publisher={American Physical Society (APS)},
   author={Pani, Paolo and Cardoso, Vitor},
   year={2009},
   month=apr }

@article{Kleihaus2011,
   title={Rotating Black Holes in Dilatonic Einstein-Gauss-Bonnet Theory},
   volume={106},
   ISSN={1079-7114},
   url={http://dx.doi.org/10.1103/PhysRevLett.106.151104},
   DOI={10.1103/physrevlett.106.151104},
   number={15},
   journal={Physical Review Letters},
   publisher={American Physical Society (APS)},
   author={Kleihaus, Burkhard and Kunz, Jutta and Radu, Eugen},
   year={2011},
   month=apr }

@article{Ayzenberg2014,
    author = "Ayzenberg, Dimitry and Yunes, Nicolas",
    title = "{Slowly-Rotating Black Holes in Einstein-Dilaton-Gauss-Bonnet Gravity: Quadratic Order in Spin Solutions}",
    eprint = "1405.2133",
    archivePrefix = "arXiv",
    primaryClass = "gr-qc",
    doi = "10.1103/PhysRevD.90.044066",
    journal = "Phys. Rev. D",
    volume = "90",
    pages = "044066",
    year = "2014",
    note = "[Erratum: Phys.Rev.D 91, 069905 (2015)]"
}

@article{Maselli2015,
  title = {Rotating black holes in Einstein-dilaton-Gauss-Bonnet gravity with finite coupling},
  author = {Maselli, Andrea and Pani, Paolo and Gualtieri, Leonardo and Ferrari, Valeria},
  journal = {Phys. Rev. D},
  volume = {92},
  issue = {8},
  pages = {083014},
  numpages = {14},
  year = {2015},
  month = {Oct},
  publisher = {American Physical Society},
  doi = {10.1103/PhysRevD.92.083014},
  url = {https://link.aps.org/doi/10.1103/PhysRevD.92.083014}
}

@article{Kleihaus2016,
  title = {Spinning black holes in Einstein--Gauss-Bonnet--dilaton theory: Nonperturbative solutions},
  author = {Kleihaus, Burkhard and Kunz, Jutta and Mojica, Sindy and Radu, Eugen},
  journal = {Phys. Rev. D},
  volume = {93},
  issue = {4},
  pages = {044047},
  numpages = {18},
  year = {2016},
  month = {Feb},
  publisher = {American Physical Society},
  doi = {10.1103/PhysRevD.93.044047},
  url = {https://link.aps.org/doi/10.1103/PhysRevD.93.044047}
}

@article{Kokkotas2017,
  title = {Analytical approximation for the Einstein-dilaton-Gauss-Bonnet black hole metric},
  author = {Kokkotas, K. D. and Konoplya, R. A. and Zhidenko, A.},
  journal = {Phys. Rev. D},
  volume = {96},
  issue = {6},
  pages = {064004},
  numpages = {8},
  year = {2017},
  month = {Sep},
  publisher = {American Physical Society},
  doi = {10.1103/PhysRevD.96.064004},
  url = {https://link.aps.org/doi/10.1103/PhysRevD.96.064004}
}

@article{AlexanderYunes2018,
  title = {Gravitational wave probes of parity violation in compact binary coalescences},
  author = {Alexander, Stephon H. and Yunes, Nicol\'as},
  journal = {Phys. Rev. D},
  volume = {97},
  issue = {6},
  pages = {064033},
  numpages = {6},
  year = {2018},
  month = {Mar},
  publisher = {American Physical Society},
  doi = {10.1103/PhysRevD.97.064033},
  url = {https://link.aps.org/doi/10.1103/PhysRevD.97.064033}
}

@article{LoutrelYunes2022,
  title = {Parity violation in spin-precessing binaries: Gravitational waves from the inspiral of black holes in dynamical Chern-Simons gravity},
  author = {Loutrel, Nicholas and Yunes, Nicol\'as},
  journal = {Phys. Rev. D},
  volume = {106},
  issue = {6},
  pages = {064009},
  numpages = {34},
  year = {2022},
  month = {Sep},
  publisher = {American Physical Society},
  doi = {10.1103/PhysRevD.106.064009},
  url = {https://link.aps.org/doi/10.1103/PhysRevD.106.064009}
}

@article{LIGO2019-1,
  title = {Tests of General Relativity with GW170817},
  author = {Abbott, B. P. and Abbott, R. and Abbott, T. D. and Acernese, F. and Ackley, K. and Adams, C. and Adams, T. and Addesso, P. and Adhikari, R. X. and Adya, V. B. and Affeldt, C. and Agarwal, B. and Agathos, M. and Agatsuma, K. and Aggarwal, N. and Aguiar, O. D. and Aiello, L. and Ain, A. and Ajith, P. and Allen, B. and Allen, G. and Allocca, A. and Aloy, M. A. and Altin, P. A. and Amato, A. and Ananyeva, A. and Anderson, S. B. and Anderson, W. G. and Angelova, S. V. and Antier, S. and Appert, S. and Arai, K. and Araya, M. C. and Areeda, J. S. and Ar\`ene, M. and Arnaud, N. and Arun, K. G. and Ascenzi, S. and Ashton, G. and Ast, M. and Aston, S. M. and Astone, P. and Atallah, D. V. and Aubin, F. and Aufmuth, P. and Aulbert, C. and AultONeal, K. and Austin, C. and Avila-Alvarez, A. and Babak, S. and Bacon, P. and Badaracco, F. and Bader, M. K. M. and Bae, S. and Baker, P. T. and Baldaccini, F. and Ballardin, G. and Ballmer, S. W. and Banagiri, S. and Barayoga, J. C. and Barclay, S. E. and Barish, B. C. and Barker, D. and Barkett, K. and Barnum, S. and Barone, F. and Barr, B. and Barsotti, L. and Barsuglia, M. and Barta, D. and Bartlett, J. and Bartos, I. and Bassiri, R. and Basti, A. and Batch, J. C. and Bawaj, M. and Bayley, J. C. and Bazzan, M. and B\'ecsy, B. and Beer, C. and Bejger, M. and Belahcene, I. and Bell, A. S. and Beniwal, D. and Bensch, M. and Berger, B. K. and Bergmann, G. and Bernuzzi, S. and Bero, J. J. and Berry, C. P. L. and Bersanetti, D. and Bertolini, A. and Betzwieser, J. and Bhandare, R. and Bilenko, I. A. and Bilgili, S. A. and Billingsley, G. and Billman, C. R. and Birch, J. and Birney, R. and Birnholtz, O. and Biscans, S. and Biscoveanu, S. and Bisht, A. and Bitossi, M. and Bizouard, M. A. and Blackburn, J. K. and Blackman, J. and Blair, C. D. and Blair, D. G. and Blair, R. M. and Bloemen, S. and Bode, N. and Boer, M. and Boetzel, Y. and Bogaert, G. and Bohe, A. and Bondu, F. and Bonilla, E. and Bonnand, R. and Booker, P. and Boom, B. A. and Booth, C. D. and Bork, R. and Boschi, V. and Bose, S. and Bossie, K. and Bossilkov, V. and Bosveld, J. and Bouffanais, Y. and Bozzi, A. and Bradaschia, C. and Brady, P. R. and Bramley, A. and Branchesi, M. and Brau, J. E. and Briant, T. and Brighenti, F. and Brillet, A. and Brinkmann, M. and Brisson, V. and Brockill, P. and Brooks, A. F. and Brown, D. D. and Brunett, S. and Buchanan, C. C. and Buikema, A. and Bulik, T. and Bulten, H. J. and Buonanno, A. and Buskulic, D. and Buy, C. and Byer, R. L. and Cabero, M. and Cadonati, L. and Cagnoli, G. and Cahillane, C. and Bustillo, J. Calder\'on and Callister, T. A. and Calloni, E. and Camp, J. B. and Canepa, M. and Canizares, P. and Cannon, K. C. and Cao, H. and Cao, J. and Capano, C. D. and Capocasa, E. and Carbognani, F. and Caride, S. and Carney, M. F. and Carullo, G. and Diaz, J. Casanueva and Casentini, C. and Caudill, S. and Cavagli\`a, M. and Cavalier, F. and Cavalieri, R. and Cella, G. and Cepeda, C. B. and Cerd\'a-Dur\'an, P. and Cerretani, G. and Cesarini, E. and Chaibi, O. and Chamberlin, S. J. and Chan, M. and Chao, S. and Charlton, P. and Chase, E. and Chassande-Mottin, E. and Chatterjee, D. and Chatziioannou, K. and Cheeseboro, B. D. and Chen, H. Y. and Chen, X. and Chen, Y. and Cheng, H.-P. and Chia, H. Y. and Chincarini, A. and Chiummo, A. and Chmiel, T. and Cho, H. S. and Cho, M. and Chow, J. H. and Christensen, N. and Chu, Q. and Chua, A. J. K. and Chua, S. and Chung, K. W. and Chung, S. and Ciani, G. and Ciobanu, A. A. and Ciolfi, R. and Cipriano, F. and Cirelli, C. E. and Cirone, A. and Clara, F. and Clark, J. A. and Clearwater, P. and Cleva, F. and Cocchieri, C. and Coccia, E. and Cohadon, P.-F. and Cohen, D. and Colla, A. and Collette, C. G. and Collins, C. and Cominsky, L. R. and Constancio, M. and Conti, L. and Cooper, S. J. and Corban, P. and Corbitt, T. R. and Cordero-Carri\'on, I. and Corley, K. R. and Cornish, N. and Corsi, A. and Cortese, S. and Costa, C. A. and Cotesta, R. and Coughlin, M. W. and Coughlin, S. B. and Coulon, J.-P. and Countryman, S. T. and Couvares, P. and Covas, P. B. and Cowan, E. E. and Coward, D. M. and Cowart, M. J. and Coyne, D. C. and Coyne, R. and Creighton, J. D. E. and Creighton, T. D. and Cripe, J. and Crowder, S. G. and Cullen, T. J. and Cumming, A. and Cunningham, L. and Cuoco, E. and Canton, T. Dal and D\'alya, G. and Danilishin, S. L. and D'Antonio, S. and Danzmann, K. and Dasgupta, A. and Costa, C. F. Da Silva and Dattilo, V. and Dave, I. and Davier, M. and Davis, D. and Daw, E. J. and Day, B. and DeBra, D. and Deenadayalan, M. and Degallaix, J. and De Laurentis, M. and Del\'eglise, S. and Del Pozzo, W. and Demos, N. and Denker, T. and Dent, T. and De Pietri, R. and Derby, J. and Dergachev, V. and De Rosa, R. and De Rossi, C. and DeSalvo, R. and de Varona, O. and Dhurandhar, S. and D\'{\i}az, M. C. and Dietrich, T. and Di Fiore, L. and Di Giovanni, M. and Di Girolamo, T. and Di Lieto, A. and Ding, B. and Di Pace, S. and Di Palma, I. and Di Renzo, F. and Dmitriev, A. and Doctor, Z. and Dolique, V. and Donovan, F. and Dooley, K. L. and Doravari, S. and Dorrington, I. and \'Alvarez, M. Dovale and Downes, T. P. and Drago, M. and Dreissigacker, C. and Driggers, J. C. and Du, Z. and Dupej, P. and Dwyer, S. E. and Easter, P. J. and Edo, T. B. and Edwards, M. C. and Effler, A. and Ehrens, P. and Eichholz, J. and Eikenberry, S. S. and Eisenmann, M. and Eisenstein, R. A. and Essick, R. C. and Estelles, H. and Estevez, D. and Etienne, Z. B. and Etzel, T. and Evans, M. and Evans, T. M. and Fafone, V. and Fair, H. and Fairhurst, S. and Fan, X. and Farinon, S. and Farr, B. and Farr, W. M. and Fauchon-Jones, E. J. and Favata, M. and Fays, M. and Fee, C. and Fehrmann, H. and Feicht, J. and Fejer, M. M. and Feng, F. and Fernandez-Galiana, A. and Ferrante, I. and Ferreira, E. C. and Ferrini, F. and Fidecaro, F. and Fiori, I. and Fiorucci, D. and Fishbach, M. and Fisher, R. P. and Fishner, J. M. and Fitz-Axen, M. and Flaminio, R. and Fletcher, M. and Fong, H. and Font, J. A. and Forsyth, P. W. F. and Forsyth, S. S. and Fournier, J.-D. and Frasca, S. and Frasconi, F. and Frei, Z. and Freise, A. and Frey, R. and Frey, V. and Fritschel, P. and Frolov, V. V. and Fulda, P. and Fyffe, M. and Gabbard, H. A. and Gadre, B. U. and Gaebel, S. M. and Gair, J. R. and Gammaitoni, L. and Ganija, M. R. and Gaonkar, S. G. and Garcia, A. and Garc\'{\i}a-Quir\'os, C. and Garufi, F. and Gateley, B. and Gaudio, S. and Gaur, G. and Gayathri, V. and Gemme, G. and Genin, E. and Gennai, A. and George, D. and George, J. and Gergely, L. and Germain, V. and Ghonge, S. and Ghosh, Abhirup and Ghosh, Archisman and Ghosh, S. and Giacomazzo, B. and Giaime, J. A. and Giardina, K. D. and Giazotto, A. and Gill, K. and Giordano, G. and Glover, L. and Goetz, E. and Goetz, R. and Goncharov, B. and Gonz\'alez, G. and Castro, J. M. Gonzalez and Gopakumar, A. and Gorodetsky, M. L. and Gossan, S. E. and Gosselin, M. and Gouaty, R. and Grado, A. and Graef, C. and Granata, M. and Grant, A. and Gras, S. and Gray, C. and Greco, G. and Green, A. C. and Green, R. and Gretarsson, E. M. and Groot, P. and Grote, H. and Grunewald, S. and Gruning, P. and Guidi, G. M. and Gulati, H. K. and Guo, X. and Gupta, A. and Gupta, M. K. and Gushwa, K. E. and Gustafson, E. K. and Gustafson, R. and Halim, O. and Hall, B. R. and Hall, E. D. and Hamilton, E. Z. and Hamilton, H. F. and Hammond, G. and Haney, M. and Hanke, M. M. and Hanks, J. and Hanna, C. and Hannam, M. D. and Hannuksela, O. A. and Hanson, J. and Hardwick, T. and Harms, J. and Harry, G. M. and Harry, I. W. and Hart, M. J. and Haster, C.-J. and Haughian, K. and Healy, J. and Heidmann, A. and Heintze, M. C. and Heitmann, H. and Hello, P. and Hemming, G. and Hendry, M. and Heng, I. S. and Hennig, J. and Heptonstall, A. W. and Hernandez, F. J. and Heurs, M. and Hild, S. and Hinderer, T. and Hoak, D. and Hochheim, S. and Hofman, D. and Holland, N. A. and Holt, K. and Holz, D. E. and Hopkins, P. and Horst, C. and Hough, J. and Houston, E. A. and Howell, E. J. and Hreibi, A. and Huerta, E. A. and Huet, D. and Hughey, B. and Hulko, M. and Husa, S. and Huttner, S. H. and Huynh-Dinh, T. and Iess, A. and Indik, N. and Ingram, C. and Inta, R. and Intini, G. and Isa, H. N. and Isac, J.-M. and Isi, M. and Iyer, B. R. and Izumi, K. and Jacqmin, T. and Jani, K. and Jaranowski, P. and Johnson, D. S. and Johnson, W. W. and Jones, D. I. and Jones, R. and Jonker, R. J. G. and Ju, L. and Junker, J. and Kalaghatgi, C. V. and Kalogera, V. and Kamai, B. and Kandhasamy, S. and Kang, G. and Kanner, J. B. and Kapadia, S. J. and Karki, S. and Karvinen, K. S. and Kasprzack, M. and Katolik, M. and Katsanevas, S. and Katsavounidis, E. and Katzman, W. and Kaufer, S. and Kawabe, K. and Keerthana, N. V. and K\'ef\'elian, F. and Keitel, D. and Kemball, A. J. and Kennedy, R. and Key, J. S. and Khalili, F. Y. and Khamesra, B. and Khan, H. and Khan, I. and Khan, S. and Khan, Z. and Khazanov, E. A. and Kijbunchoo, N. and Kim, Chunglee and Kim, J. C. and Kim, K. and Kim, W. and Kim, W. S. and Kim, Y.-M. and King, E. J. and King, P. J. and Kinley-Hanlon, M. and Kirchhoff, R. and Kissel, J. S. and Kleybolte, L. and Klimenko, S. and Knowles, T. D. and Koch, P. and Koehlenbeck, S. M. and Koley, S. and Kondrashov, V. and Kontos, A. and Korobko, M. and Korth, W. Z. and Kowalska, I. and Kozak, D. B. and Kr\"amer, C. and Kringel, V. and Krishnan, B. and Kr\'olak, A. and Kuehn, G. and Kumar, P. and Kumar, R. and Kumar, S. and Kuo, L. and Kutynia, A. and Kwang, S. and Lackey, B. D. and Lai, K. H. and Landry, M. and Lang, R. N. and Lange, J. and Lantz, B. and Lanza, R. K. and Lartaux-Vollard, A. and Lasky, P. D. and Laxen, M. and Lazzarini, A. and Lazzaro, C. and Leaci, P. and Leavey, S. and Lee, C. H. and Lee, H. K. and Lee, H. M. and Lee, H. W. and Lee, K. and Lehmann, J. and Lenon, A. and Leonardi, M. and Leroy, N. and Letendre, N. and Levin, Y. and Li, J. and Li, T. G. F. and Li, X. and Linker, S. D. and Littenberg, T. B. and Liu, J. and Liu, X. and Lo, R. K. L. and Lockerbie, N. A. and London, L. T. and Longo, A. and Lorenzini, M. and Loriette, V. and Lormand, M. and Losurdo, G. and Lough, J. D. and Lousto, C. O. and Lovelace, G. and L\"uck, H. and Lumaca, D. and Lundgren, A. P. and Lynch, R. and Ma, Y. and Macas, R. and Macfoy, S. and Machenschalk, B. and MacInnis, M. and Macleod, D. M. and Hernandez, I. Maga\~na and Maga\~na-Sandoval, F. and Zertuche, L. Maga\~na and Magee, R. M. and Majorana, E. and Maksimovic, I. and Man, N. and Mandic, V. and Mangano, V. and Mansell, G. L. and Manske, M. and Mantovani, M. and Marchesoni, F. and Marion, F. and M\'arka, S. and M\'arka, Z. and Markakis, C. and Markosyan, A. S. and Markowitz, A. and Maros, E. and Marquina, A. and Marsat, S. and Martelli, F. and Martellini, L. and Martin, I. W. and Martin, R. M. and Martynov, D. V. and Mason, K. and Massera, E. and Masserot, A. and Massinger, T. J. and Masso-Reid, M. and Mastrogiovanni, S. and Matas, A. and Matichard, F. and Matone, L. and Mavalvala, N. and Mazumder, N. and McCann, J. J. and McCarthy, R. and McClelland, D. E. and McCormick, S. and McCuller, L. and McGuire, S. C. and McIver, J. and McManus, D. J. and McRae, T. and McWilliams, S. T. and Meacher, D. and Meadors, G. D. and Mehmet, M. and Meidam, J. and Mejuto-Villa, E. and Melatos, A. and Mendell, G. and Mendoza-Gandara, D. and Mercer, R. A. and Mereni, L. and Merilh, E. L. and Merzougui, M. and Meshkov, S. and Messenger, C. and Messick, C. and Metzdorff, R. and Meyers, P. M. and Miao, H. and Michel, C. and Middleton, H. and Mikhailov, E. E. and Milano, L. and Miller, A. L. and Miller, A. and Miller, B. B. and Miller, J. and Millhouse, M. and Mills, J. and Milovich-Goff, M. C. and Minazzoli, O. and Minenkov, Y. and Ming, J. and Mishra, C. and Mitra, S. and Mitrofanov, V. P. and Mitselmakher, G. and Mittleman, R. and Moffa, D. and Mogushi, K. and Mohan, M. and Mohapatra, S. R. P. and Montani, M. and Moore, C. J. and Moraru, D. and Moreno, G. and Morisaki, S. and Mours, B. and Mow-Lowry, C. M. and Mueller, G. and Muir, A. W. and Mukherjee, Arunava and Mukherjee, D. and Mukherjee, S. and Mukund, N. and Mullavey, A. and Munch, J. and Mu\~niz, E. A. and Muratore, M. and Murray, P. G. and Nagar, A. and Napier, K. and Nardecchia, I. and Naticchioni, L. and Nayak, R. K. and Neilson, J. and Nelemans, G. and Nelson, T. J. N. and Nery, M. and Neunzert, A. and Nevin, L. and Newport, J. M. and Ng, K. Y. and Ng, S. and Nguyen, P. and Nguyen, T. T. and Nichols, D. and Nielsen, A. B. and Nissanke, S. and Nitz, A. and Nocera, F. and Nolting, D. and North, C. and Nuttall, L. K. and Obergaulinger, M. and Oberling, J. and O'Brien, B. D. and O'Dea, G. D. and Ogin, G. H. and Oh, J. J. and Oh, S. H. and Ohme, F. and Ohta, H. and Okada, M. A. and Oliver, M. and Oppermann, P. and Oram, Richard J. and O'Reilly, B. and Ormiston, R. and Ortega, L. F. and O'Shaughnessy, R. and Ossokine, S. and Ottaway, D. J. and Overmier, H. and Owen, B. J. and Pace, A. E. and Pagano, G. and Page, J. and Page, M. A. and Pai, A. and Pai, S. A. and Palamos, J. R. and Palashov, O. and Palomba, C. and Pal-Singh, A. and Pan, Howard and Pan, Huang-Wei and Pang, B. and Pang, P. T. H. and Pankow, C. and Pannarale, F. and Pant, B. C. and Paoletti, F. and Paoli, A. and Papa, M. A. and Parida, A. and Parker, W. and Pascucci, D. and Pasqualetti, A. and Passaquieti, R. and Passuello, D. and Patil, M. and Patricelli, B. and Pearlstone, B. L. and Pedersen, C. and Pedraza, M. and Pedurand, R. and Pekowsky, L. and Pele, A. and Penn, S. and Perez, C. J. and Perreca, A. and Perri, L. M. and Pfeiffer, H. P. and Phelps, M. and Phukon, K. S. and Piccinni, O. J. and Pichot, M. and Piergiovanni, F. and Pierro, V. and Pillant, G. and Pinard, L. and Pinto, I. M. and Pirello, M. and Pitkin, M. and Poggiani, R. and Popolizio, P. and Porter, E. K. and Possenti, L. and Post, A. and Powell, J. and Prasad, J. and Pratt, J. W. W. and Pratten, G. and Predoi, V. and Prestegard, T. and Principe, M. and Privitera, S. and Prodi, G. A. and Prokhorov, L. G. and Puncken, O. and Punturo, M. and Puppo, P. and P\"urrer, M. and Qi, H. and Quetschke, V. and Quintero, E. A. and Quitzow-James, R. and Raab, F. J. and Rabeling, D. S. and Radkins, H. and Raffai, P. and Raja, S. and Rajan, C. and Rajbhandari, B. and Rakhmanov, M. and Ramirez, K. E. and Ramos-Buades, A. and Rana, Javed and Rapagnani, P. and Raymond, V. and Razzano, M. and Read, J. and Regimbau, T. and Rei, L. and Reid, S. and Reitze, D. H. and Ren, W. and Ricci, F. and Ricker, P. M. and Riemenschneider, G. M. and Riles, K. and Rizzo, M. and Robertson, N. A. and Robie, R. and Robinet, F. and Robson, T. and Rocchi, A. and Rolland, L. and Rollins, J. G. and Roma, V. J. and Romano, R. and Romel, C. L. and Romie, J. H. and Rosi\ifmmode \acute{n}\else \'{n}\fi{}ska, D. and Ross, M. P. and Rowan, S. and R\"udiger, A. and Ruggi, P. and Rutins, G. and Ryan, K. and Sachdev, S. and Sadecki, T. and Sakellariadou, M. and Salconi, L. and Saleem, M. and Salemi, F. and Samajdar, A. and Sammut, L. and Sampson, L. M. and Sanchez, E. J. and Sanchez, L. E. and Sanchis-Gual, N. and Sandberg, V. and Sanders, J. R. and Sarin, N. and Sassolas, B. and Sathyaprakash, B. S. and Saulson, P. R. and Sauter, O. and Savage, R. L. and Sawadsky, A. and Schale, P. and Scheel, M. and Scheuer, J. and Schmidt, P. and Schnabel, R. and Schofield, R. M. S. and Sch\"onbeck, A. and Schreiber, E. and Schuette, D. and Schulte, B. W. and Schutz, B. F. and Schwalbe, S. G. and Scott, J. and Scott, S. M. and Seidel, E. and Sellers, D. and Sengupta, A. S. and Sennett, N. and Sentenac, D. and Sequino, V. and Sergeev, A. and Setyawati, Y. and Shaddock, D. A. and Shaffer, T. J. and Shah, A. A. and Shahriar, M. S. and Shaner, M. B. and Shao, L. and Shapiro, B. and Shawhan, P. and Shen, H. and Shoemaker, D. H. and Shoemaker, D. M. and Siellez, K. and Siemens, X. and Sieniawska, M. and Sigg, D. and Silva, A. D. and Singer, L. P. and Singh, A. and Singhal, A. and Sintes, A. M. and Slagmolen, B. J. J. and Slaven-Blair, T. J. and Smith, B. and Smith, J. R. and Smith, R. J. E. and Somala, S. and Son, E. J. and Sorazu, B. and Sorrentino, F. and Souradeep, T. and Spencer, A. P. and Srivastava, A. K. and Staats, K. and Steer, D. A. and Steinke, M. and Steinlechner, J. and Steinlechner, S. and Steinmeyer, D. and Steltner, B. and Stevenson, S. P. and Stocks, D. and Stone, R. and Stops, D. J. and Strain, K. A. and Stratta, G. and Strigin, S. E. and Strunk, A. and Sturani, R. and Stuver, A. L. and Summerscales, T. Z. and Sun, L. and Sunil, S. and Suresh, J. and Sutton, P. J. and Swinkels, B. L. and Szczepa\ifmmode \acute{n}\else \'{n}\fi{}czyk, M. J. and Tacca, M. and Tait, S. C. and Talbot, C. and Talukder, D. and Tamanini, N. and Tanner, D. B. and T\'apai, M. and Taracchini, A. and Tasson, J. D. and Taylor, J. A. and Taylor, R. and Tewari, S. V. and Theeg, T. and Thies, F. and Thomas, E. G. and Thomas, M. and Thomas, P. and Thorne, K. A. and Thrane, E. and Tiwari, S. and Tiwari, V. and Tokmakov, K. V. and Toland, K. and Tonelli, M. and Tornasi, Z. and Torres-Forn\'e, A. and Torrie, C. I. and T\"oyr\"a, D. and Travasso, F. and Traylor, G. and Trinastic, J. and Tringali, M. C. and Trozzo, L. and Tsang, K. W. and Tse, M. and Tso, R. and Tsukada, L. and Tsuna, D. and Tuyenbayev, D. and Ueno, K. and Ugolini, D. and Urban, A. L. and Usman, S. A. and Vahlbruch, H. and Vajente, G. and Valdes, G. and van Bakel, N. and van Beuzekom, M. and van den Brand, J. F. J. and Van Den Broeck, C. and Vander-Hyde, D. C. and van der Schaaf, L. and van Heijningen, J. V. and van Veggel, A. A. and Vardaro, M. and Varma, V. and Vass, S. and Vas\'uth, M. and Vecchio, A. and Vedovato, G. and Veitch, J. and Veitch, P. J. and Venkateswara, K. and Venugopalan, G. and Verkindt, D. and Vetrano, F. and Vicer\'e, A. and Viets, A. D. and Vinciguerra, S. and Vine, D. J. and Vinet, J.-Y. and Vitale, S. and Vo, T. and Vocca, H. and Vorvick, C. and Vyatchanin, S. P. and Wade, A. R. and Wade, L. E. and Wade, M. and Walet, R. and Walker, M. and Wallace, L. and Walsh, S. and Wang, G. and Wang, H. and Wang, J. Z. and Wang, W. H. and Wang, Y. F. and Ward, R. L. and Warner, J. and Was, M. and Watchi, J. and Weaver, B. and Wei, L.-W. and Weinert, M. and Weinstein, A. J. and Weiss, R. and Wellmann, F. and Wen, L. and Wessel, E. K. and We\ss{}els, P. and Westerweck, J. and Wette, K. and Whelan, J. T. and Whiting, B. F. and Whittle, C. and Wilken, D. and Williams, D. and Williams, R. D. and Williamson, A. R. and Willis, J. L. and Willke, B. and Wimmer, M. H. and Winkler, W. and Wipf, C. C. and Wittel, H. and Woan, G. and Woehler, J. and Wofford, J. K. and Wong, W. K. and Worden, J. and Wright, J. L. and Wu, D. S. and Wysocki, D. M. and Xiao, S. and Yam, W. and Yamamoto, H. and Yancey, C. C. and Yang, L. and Yap, M. J. and Yazback, M. and Yu, Hang and Yu, Haocun and Yvert, M. and Zadro\ifmmode \dot{z}\else \.{z}\fi{}ny, A. and Zanolin, M. and Zelenova, T. and Zendri, J.-P. and Zevin, M. and Zhang, J. and Zhang, L. and Zhang, M. and Zhang, T. and Zhang, Y.-H. and Zhao, C. and Zhou, M. and Zhou, Z. and Zhu, S. J. and Zhu, X. J. and Zimmerman, A. B. and Zucker, M. E. and Zweizig, J.},
  collaboration = {LIGO Scientific Collaboration and Virgo Collaboration},
  journal = {Phys. Rev. Lett.},
  volume = {123},
  issue = {1},
  pages = {011102},
  numpages = {15},
  year = {2019},
  month = {Jul},
  publisher = {American Physical Society},
  doi = {10.1103/PhysRevLett.123.011102},
  url = {https://link.aps.org/doi/10.1103/PhysRevLett.123.011102}
}

@article{LIGO2019-2,
  title = {Tests of general relativity with the binary black hole signals from the LIGO-Virgo catalog GWTC-1},
  author = {Abbott, B. P. and Abbott, R. and Abbott, T. D. and Abraham, S. and Acernese, F. and Ackley, K. and Adams, C. and Adhikari, R. X. and Adya, V. B. and Affeldt, C. and Agathos, M. and Agatsuma, K. and Aggarwal, N. and Aguiar, O. D. and Aiello, L. and Ain, A. and Ajith, P. and Allen, G. and Allocca, A. and Aloy, M. A. and Altin, P. A. and Amato, A. and Ananyeva, A. and Anderson, S. B. and Anderson, W. G. and Angelova, S. V. and Antier, S. and Appert, S. and Arai, K. and Araya, M. C. and Areeda, J. S. and Ar\`ene, M. and Arnaud, N. and Arun, K. G. and Ascenzi, S. and Ashton, G. and Aston, S. M. and Astone, P. and Aubin, F. and Aufmuth, P. and AultONeal, K. and Austin, C. and Avendano, V. and Avila-Alvarez, A. and Babak, S. and Bacon, P. and Badaracco, F. and Bader, M. K. M. and Bae, S. and Baker, P. T. and Baldaccini, F. and Ballardin, G. and Ballmer, S. W. and Banagiri, S. and Barayoga, J. C. and Barclay, S. E. and Barish, B. C. and Barker, D. and Barkett, K. and Barnum, S. and Barone, F. and Barr, B. and Barsotti, L. and Barsuglia, M. and Barta, D. and Bartlett, J. and Bartos, I. and Bassiri, R. and Basti, A. and Bawaj, M. and Bayley, J. C. and Bazzan, M. and B\'ecsy, B. and Bejger, M. and Belahcene, I. and Bell, A. S. and Beniwal, D. and Berger, B. K. and Bergmann, G. and Bernuzzi, S. and Bero, J. J. and Berry, C. P. L. and Bersanetti, D. and Bertolini, A. and Betzwieser, J. and Bhandare, R. and Bidler, J. and Bilenko, I. A. and Bilgili, S. A. and Billingsley, G. and Birch, J. and Birney, R. and Birnholtz, O. and Biscans, S. and Biscoveanu, S. and Bisht, A. and Bitossi, M. and Bizouard, M. A. and Blackburn, J. K. and Blair, C. D. and Blair, D. G. and Blair, R. M. and Bloemen, S. and Bode, N. and Boer, M. and Boetzel, Y. and Bogaert, G. and Bondu, F. and Bonilla, E. and Bonnand, R. and Booker, P. and Boom, B. A. and Booth, C. D. and Bork, R. and Boschi, V. and Bose, S. and Bossie, K. and Bossilkov, V. and Bosveld, J. and Bouffanais, Y. and Bozzi, A. and Bradaschia, C. and Brady, P. R. and Bramley, A. and Branchesi, M. and Brau, J. E. and Breschi, M. and Briant, T. and Briggs, J. H. and Brighenti, F. and Brillet, A. and Brinkmann, M. and Brisson, V. and Brito, R. and Brockill, P. and Brooks, A. F. and Brown, D. D. and Brunett, S. and Buikema, A. and Bulik, T. and Bulten, H. J. and Buonanno, A. and Buskulic, D. and Rosell, M. J. Bustamante and Buy, C. and Byer, R. L. and Cabero, M. and Cadonati, L. and Cagnoli, G. and Cahillane, C. and Bustillo, J. Calder\'on and Callister, T. A. and Calloni, E. and Camp, J. B. and Campbell, W. A. and Canepa, M. and Cannon, K. C. and Cao, H. and Cao, J. and Capano, C. D. and Capocasa, E. and Carbognani, F. and Caride, S. and Carney, M. F. and Carullo, G. and Diaz, J. Casanueva and Casentini, C. and Caudill, S. and Cavagli\`a, M. and Cavalier, F. and Cavalieri, R. and Cella, G. and Cerd\'a-Dur\'an, P. and Cerretani, G. and Cesarini, E. and Chaibi, O. and Chakravarti, K. and Chamberlin, S. J. and Chan, M. and Chao, S. and Charlton, P. and Chase, E. A. and Chassande-Mottin, E. and Chatterjee, D. and Chaturvedi, M. and Chatziioannou, K. and Cheeseboro, B. D. and Chen, H. Y. and Chen, X. and Chen, Y. and Cheng, H.-P. and Cheong, C. K. and Chia, H. Y. and Chincarini, A. and Chiummo, A. and Cho, G. and Cho, H. S. and Cho, M. and Christensen, N. and Chu, Q. and Chua, S. and Chung, K. W. and Chung, S. and Ciani, G. and Ciobanu, A. A. and Ciolfi, R. and Cipriano, F. and Cirone, A. and Clara, F. and Clark, J. A. and Clearwater, P. and Cleva, F. and Cocchieri, C. and Coccia, E. and Cohadon, P.-F. and Cohen, D. and Colgan, R. and Colleoni, M. and Collette, C. G. and Collins, C. and Cominsky, L. R. and Constancio, M. and Conti, L. and Cooper, S. J. and Corban, P. and Corbitt, T. R. and Cordero-Carri\'on, I. and Corley, K. R. and Cornish, N. and Corsi, A. and Cortese, S. and Costa, C. A. and Cotesta, R. and Coughlin, M. W. and Coughlin, S. B. and Coulon, J.-P. and Countryman, S. T. and Couvares, P. and Covas, P. B. and Cowan, E. E. and Coward, D. M. and Cowart, M. J. and Coyne, D. C. and Coyne, R. and Creighton, J. D. E. and Creighton, T. D. and Cripe, J. and Croquette, M. and Crowder, S. G. and Cullen, T. J. and Cumming, A. and Cunningham, L. and Cuoco, E. and Canton, T. Dal and D\'alya, G. and Danilishin, S. L. and D'Antonio, S. and Danzmann, K. and Dasgupta, A. and Costa, C. F. Da Silva and Datrier, L. E. H. and Dattilo, V. and Dave, I. and Davier, M. and Davis, D. and Daw, E. J. and DeBra, D. and Deenadayalan, M. and Degallaix, J. and De Laurentis, M. and Del\'eglise, S. and Del Pozzo, W. and DeMarchi, L. M. and Demos, N. and Dent, T. and De Pietri, R. and Derby, J. and De Rosa, R. and De Rossi, C. and DeSalvo, R. and de Varona, O. and Dhurandhar, S. and D\'{\i}az, M. C. and Dietrich, T. and Di Fiore, L. and Di Giovanni, M. and Di Girolamo, T. and Di Lieto, A. and Ding, B. and Di Pace, S. and Di Palma, I. and Di Renzo, F. and Dmitriev, A. and Doctor, Z. and Donovan, F. and Dooley, K. L. and Doravari, S. and Dorrington, I. and Downes, T. P. and Drago, M. and Driggers, J. C. and Du, Z. and Ducoin, J.-G. and Dupej, P. and Dwyer, S. E. and Easter, P. J. and Edo, T. B. and Edwards, M. C. and Effler, A. and Ehrens, P. and Eichholz, J. and Eikenberry, S. S. and Eisenmann, M. and Eisenstein, R. A. and Essick, R. C. and Estelles, H. and Estevez, D. and Etienne, Z. B. and Etzel, T. and Evans, M. and Evans, T. M. and Fafone, V. and Fair, H. and Fairhurst, S. and Fan, X. and Farinon, S. and Farr, B. and Farr, W. M. and Fauchon-Jones, E. J. and Favata, M. and Fays, M. and Fazio, M. and Fee, C. and Feicht, J. and Fejer, M. M. and Feng, F. and Fernandez-Galiana, A. and Ferrante, I. and Ferreira, E. C. and Ferreira, T. A. and Ferrini, F. and Fidecaro, F. and Fiori, I. and Fiorucci, D. and Fishbach, M. and Fisher, R. P. and Fishner, J. M. and Fitz-Axen, M. and Flaminio, R. and Fletcher, M. and Flynn, E. and Fong, H. and Font, J. A. and Forsyth, P. W. F. and Fournier, J.-D. and Frasca, S. and Frasconi, F. and Frei, Z. and Freise, A. and Frey, R. and Frey, V. and Fritschel, P. and Frolov, V. V. and Fulda, P. and Fyffe, M. and Gabbard, H. A. and Gadre, B. U. and Gaebel, S. M. and Gair, J. R. and Gammaitoni, L. and Ganija, M. R. and Gaonkar, S. G. and Garcia, A. and Garc\'{\i}a-Quir\'os, C. and Garufi, F. and Gateley, B. and Gaudio, S. and Gaur, G. and Gayathri, V. and Gemme, G. and Genin, E. and Gennai, A. and George, D. and George, J. and Gergely, L. and Germain, V. and Ghonge, S. and Ghosh, Abhirup and Ghosh, Archisman and Ghosh, S. and Giacomazzo, B. and Giaime, J. A. and Giardina, K. D. and Giazotto, A. and Gill, K. and Giordano, G. and Glover, L. and Godwin, P. and Goetz, E. and Goetz, R. and Goncharov, B. and Gonz\'alez, G. and Castro, J. M. Gonzalez and Gopakumar, A. and Gorodetsky, M. L. and Gossan, S. E. and Gosselin, M. and Gouaty, R. and Grado, A. and Graef, C. and Granata, M. and Grant, A. and Gras, S. and Grassia, P. and Gray, C. and Gray, R. and Greco, G. and Green, A. C. and Green, R. and Gretarsson, E. M. and Groot, P. and Grote, H. and Grunewald, S. and Gruning, P. and Guidi, G. M. and Gulati, H. K. and Guo, Y. and Gupta, A. and Gupta, M. K. and Gustafson, E. K. and Gustafson, R. and Haegel, L. and Halim, O. and Hall, B. R. and Hall, E. D. and Hamilton, E. Z. and Hammond, G. and Haney, M. and Hanke, M. M. and Hanks, J. and Hanna, C. and Hannam, M. D. and Hannuksela, O. A. and Hanson, J. and Hardwick, T. and Haris, K. and Harms, J. and Harry, G. M. and Harry, I. W. and Haster, C.-J. and Haughian, K. and Hayes, F. J. and Healy, J. and Heidmann, A. and Heintze, M. C. and Heitmann, H. and Hello, P. and Hemming, G. and Hendry, M. and Heng, I. S. and Hennig, J. and Heptonstall, A. W. and Vivanco, Francisco Hernandez and Heurs, M. and Hild, S. and Hinderer, T. and Hoak, D. and Hochheim, S. and Hofman, D. and Holgado, A. M. and Holland, N. A. and Holt, K. and Holz, D. E. and Hopkins, P. and Horst, C. and Hough, J. and Howell, E. J. and Hoy, C. G. and Hreibi, A. and Huerta, E. A. and Huet, D. and Hughey, B. and Hulko, M. and Husa, S. and Huttner, S. H. and Huynh-Dinh, T. and Idzkowski, B. and Iess, A. and Ingram, C. and Inta, R. and Intini, G. and Irwin, B. and Isa, H. N. and Isac, J.-M. and Isi, M. and Iyer, B. R. and Izumi, K. and Jacqmin, T. and Jadhav, S. J. and Jani, K. and Janthalur, N. N. and Jaranowski, P. and Jenkins, A. C. and Jiang, J. and Johnson, D. S. and Johnson-McDaniel, N. K. and Jones, A. W. and Jones, D. I. and Jones, R. and Jonker, R. J. G. and Ju, L. and Junker, J. and Kalaghatgi, C. V. and Kalogera, V. and Kamai, B. and Kandhasamy, S. and Kang, G. and Kanner, J. B. and Kapadia, S. J. and Karki, S. and Karvinen, K. S. and Kashyap, R. and Kasprzack, M. and Katsanevas, S. and Katsavounidis, E. and Katzman, W. and Kaufer, S. and Kawabe, K. and Keerthana, N. V. and K\'ef\'elian, F. and Keitel, D. and Kennedy, R. and Key, J. S. and Khalili, F. Y. and Khan, H. and Khan, I. and Khan, S. and Khan, Z. and Khazanov, E. A. and Khursheed, M. and Kijbunchoo, N. and Kim, Chunglee and Kim, J. C. and Kim, K. and Kim, W. and Kim, W. S. and Kim, Y.-M. and Kimball, C. and King, E. J. and King, P. J. and Kinley-Hanlon, M. and Kirchhoff, R. and Kissel, J. S. and Kleybolte, L. and Klika, J. H. and Klimenko, S. and Knowles, T. D. and Koch, P. and Koehlenbeck, S. M. and Koekoek, G. and Koley, S. and Kondrashov, V. and Kontos, A. and Koper, N. and Korobko, M. and Korth, W. Z. and Kowalska, I. and Kozak, D. B. and Kringel, V. and Krishnendu, N. and Kr\'olak, A. and Kuehn, G. and Kumar, A. and Kumar, P. and Kumar, R. and Kumar, S. and Kuo, L. and Kutynia, A. and Kwang, S. and Lackey, B. D. and Lai, K. H. and Lam, T. L. and Landry, M. and Lane, B. B. and Lang, R. N. and Lange, J. and Lantz, B. and Lanza, R. K. and Lartaux-Vollard, A. and Lasky, P. D. and Laxen, M. and Lazzarini, A. and Lazzaro, C. and Leaci, P. and Leavey, S. and Lecoeuche, Y. K. and Lee, C. H. and Lee, H. K. and Lee, H. M. and Lee, H. W. and Lee, J. and Lee, K. and Lehmann, J. and Lenon, A. and Leroy, N. and Letendre, N. and Levin, Y. and Li, J. and Li, K. J. L. and Li, T. G. F. and Li, X. and Lin, F. and Linde, F. and Linker, S. D. and Littenberg, T. B. and Liu, J. and Liu, X. and Lo, R. K. L. and Lockerbie, N. A. and London, L. T. and Longo, A. and Lorenzini, M. and Loriette, V. and Lormand, M. and Losurdo, G. and Lough, J. D. and Lousto, C. O. and Lovelace, G. and Lower, M. E. and L\"uck, H. and Lumaca, D. and Lundgren, A. P. and Lynch, R. and Ma, Y. and Macas, R. and Macfoy, S. and MacInnis, M. and Macleod, D. M. and Macquet, A. and Maga\~na-Sandoval, F. and Zertuche, L. Maga\~na and Magee, R. M. and Majorana, E. and Maksimovic, I. and Malik, A. and Man, N. and Mandic, V. and Mangano, V. and Mansell, G. L. and Manske, M. and Mantovani, M. and Marchesoni, F. and Marion, F. and M\'arka, S. and M\'arka, Z. and Markakis, C. and Markosyan, A. S. and Markowitz, A. and Maros, E. and Marquina, A. and Marsat, S. and Martelli, F. and Martin, I. W. and Martin, R. M. and Martynov, D. V. and Mason, K. and Massera, E. and Masserot, A. and Massinger, T. J. and Masso-Reid, M. and Mastrogiovanni, S. and Matas, A. and Matichard, F. and Matone, L. and Mavalvala, N. and Mazumder, N. and McCann, J. J. and McCarthy, R. and McClelland, D. E. and McCormick, S. and McCuller, L. and McGuire, S. C. and McIver, J. and McManus, D. J. and McRae, T. and McWilliams, S. T. and Meacher, D. and Meadors, G. D. and Mehmet, M. and Mehta, A. K. and Meidam, J. and Melatos, A. and Mendell, G. and Mercer, R. A. and Mereni, L. and Merilh, E. L. and Merzougui, M. and Meshkov, S. and Messenger, C. and Messick, C. and Metzdorff, R. and Meyers, P. M. and Miao, H. and Michel, C. and Middleton, H. and Mikhailov, E. E. and Milano, L. and Miller, A. L. and Miller, A. and Millhouse, M. and Mills, J. C. and Milovich-Goff, M. C. and Minazzoli, O. and Minenkov, Y. and Mishkin, A. and Mishra, C. and Mistry, T. and Mitra, S. and Mitrofanov, V. P. and Mitselmakher, G. and Mittleman, R. and Mo, G. and Moffa, D. and Mogushi, K. and Mohapatra, S. R. P. and Montani, M. and Moore, C. J. and Moraru, D. and Moreno, G. and Morisaki, S. and Mours, B. and Mow-Lowry, C. M. and Mukherjee, Arunava and Mukherjee, D. and Mukherjee, S. and Mukund, N. and Mullavey, A. and Munch, J. and Mu\~niz, E. A. and Muratore, M. and Murray, P. G. and Nagar, A. and Nardecchia, I. and Naticchioni, L. and Nayak, R. K. and Neilson, J. and Nelemans, G. and Nelson, T. J. N. and Nery, M. and Neunzert, A. and Ng, K. Y. and Ng, S. and Nguyen, P. and Nichols, D. and Nielsen, A. B. and Nissanke, S. and Nitz, A. and Nocera, F. and North, C. and Nuttall, L. K. and Obergaulinger, M. and Oberling, J. and O'Brien, B. D. and O'Dea, G. D. and Ogin, G. H. and Oh, J. J. and Oh, S. H. and Ohme, F. and Ohta, H. and Okada, M. A. and Oliver, M. and Oppermann, P. and Oram, Richard J. and O'Reilly, B. and Ormiston, R. G. and Ortega, L. F. and O'Shaughnessy, R. and Ossokine, S. and Ottaway, D. J. and Overmier, H. and Owen, B. J. and Pace, A. E. and Pagano, G. and Page, M. A. and Pai, A. and Pai, S. A. and Palamos, J. R. and Palashov, O. and Palomba, C. and Pal-Singh, A. and Pan, Huang-Wei and Pang, B. and Pang, P. T. H. and Pankow, C. and Pannarale, F. and Pant, B. C. and Paoletti, F. and Paoli, A. and Parida, A. and Parker, W. and Pascucci, D. and Pasqualetti, A. and Passaquieti, R. and Passuello, D. and Patil, M. and Patricelli, B. and Pearlstone, B. L. and Pedersen, C. and Pedraza, M. and Pedurand, R. and Pele, A. and Penn, S. and Perez, C. J. and Perreca, A. and Pfeiffer, H. P. and Phelps, M. and Phukon, K. S. and Piccinni, O. J. and Pichot, M. and Piergiovanni, F. and Pillant, G. and Pinard, L. and Pirello, M. and Pitkin, M. and Poggiani, R. and Pong, D. Y. T. and Ponrathnam, S. and Popolizio, P. and Porter, E. K. and Powell, J. and Prajapati, A. K. and Prasad, J. and Prasai, K. and Prasanna, R. and Pratten, G. and Prestegard, T. and Privitera, S. and Prodi, G. A. and Prokhorov, L. G. and Puncken, O. and Punturo, M. and Puppo, P. and P\"urrer, M. and Qi, H. and Quetschke, V. and Quinonez, P. J. and Quintero, E. A. and Quitzow-James, R. and Raab, F. J. and Radkins, H. and Radulescu, N. and Raffai, P. and Raja, S. and Rajan, C. and Rajbhandari, B. and Rakhmanov, M. and Ramirez, K. E. and Ramos-Buades, A. and Rana, Javed and Rao, K. and Rapagnani, P. and Raymond, V. and Razzano, M. and Read, J. and Regimbau, T. and Rei, L. and Reid, S. and Reitze, D. H. and Ren, W. and Ricci, F. and Richardson, C. J. and Richardson, J. W. and Ricker, P. M. and Riles, K. and Rizzo, M. and Robertson, N. A. and Robie, R. and Robinet, F. and Rocchi, A. and Rolland, L. and Rollins, J. G. and Roma, V. J. and Romanelli, M. and Romano, R. and Romel, C. L. and Romie, J. H. and Rose, K. and Rosi\ifmmode \acute{n}\else \'{n}\fi{}ska, D. and Rosofsky, S. G. and Ross, M. P. and Rowan, S. and R\"udiger, A. and Ruggi, P. and Rutins, G. and Ryan, K. and Sachdev, S. and Sadecki, T. and Sakellariadou, M. and Salconi, L. and Saleem, M. and Samajdar, A. and Sammut, L. and Sanchez, E. J. and Sanchez, L. E. and Sanchis-Gual, N. and Sandberg, V. and Sanders, J. R. and Santiago, K. A. and Sarin, N. and Sassolas, B. and Sathyaprakash, B. S. and Saulson, P. R. and Sauter, O. and Savage, R. L. and Schale, P. and Scheel, M. and Scheuer, J. and Schmidt, P. and Schnabel, R. and Schofield, R. M. S. and Sch\"onbeck, A. and Schreiber, E. and Schulte, B. W. and Schutz, B. F. and Schwalbe, S. G. and Scott, J. and Scott, S. M. and Seidel, E. and Sellers, D. and Sengupta, A. S. and Sennett, N. and Sentenac, D. and Sequino, V. and Sergeev, A. and Setyawati, Y. and Shaddock, D. A. and Shaffer, T. and Shahriar, M. S. and Shaner, M. B. and Shao, L. and Sharma, P. and Shawhan, P. and Shen, H. and Shink, R. and Shoemaker, D. H. and Shoemaker, D. M. and ShyamSundar, S. and Siellez, K. and Sieniawska, M. and Sigg, D. and Silva, A. D. and Singer, L. P. and Singh, N. and Singhal, A. and Sintes, A. M. and Sitmukhambetov, S. and Skliris, V. and Slagmolen, B. J. J. and Slaven-Blair, T. J. and Smith, J. R. and Smith, R. J. E. and Somala, S. and Son, E. J. and Sorazu, B. and Sorrentino, F. and Souradeep, T. and Sowell, E. and Spencer, A. P. and Srivastava, A. K. and Srivastava, V. and Staats, K. and Stachie, C. and Standke, M. and Steer, D. A. and Steinke, M. and Steinlechner, J. and Steinlechner, S. and Steinmeyer, D. and Stevenson, S. P. and Stocks, D. and Stone, R. and Stops, D. J. and Strain, K. A. and Stratta, G. and Strigin, S. E. and Strunk, A. and Sturani, R. and Stuver, A. L. and Sudhir, V. and Summerscales, T. Z. and Sun, L. and Sunil, S. and Suresh, J. and Sutton, P. J. and Swinkels, B. L. and Szczepa\ifmmode \acute{n}\else \'{n}\fi{}czyk, M. J. and Tacca, M. and Tait, S. C. and Talbot, C. and Talukder, D. and Tanner, D. B. and T\'apai, M. and Taracchini, A. and Tasson, J. D. and Taylor, R. and Thies, F. and Thomas, M. and Thomas, P. and Thondapu, S. R. and Thorne, K. A. and Thrane, E. and Tiwari, Shubhanshu and Tiwari, Srishti and Tiwari, V. and Toland, K. and Tonelli, M. and Tornasi, Z. and Torres-Forn\'e, A. and Torrie, C. I. and T\"oyr\"a, D. and Travasso, F. and Traylor, G. and Tringali, M. C. and Trovato, A. and Trozzo, L. and Trudeau, R. and Tsang, K. W. and Tse, M. and Tso, R. and Tsukada, L. and Tsuna, D. and Tuyenbayev, D. and Ueno, K. and Ugolini, D. and Unnikrishnan, C. S. and Urban, A. L. and Usman, S. A. and Vahlbruch, H. and Vajente, G. and Valdes, G. and van Bakel, N. and van Beuzekom, M. and van den Brand, J. F. J. and Van Den Broeck, C. and Vander-Hyde, D. C. and van Heijningen, J. V. and van der Schaaf, L. and van Veggel, A. A. and Vardaro, M. and Varma, V. and Vass, S. and Vas\'uth, M. and Vecchio, A. and Vedovato, G. and Veitch, J. and Veitch, P. J. and Venkateswara, K. and Venugopalan, G. and Verkindt, D. and Vetrano, F. and Vicer\'e, A. and Viets, A. D. and Vine, D. J. and Vinet, J.-Y. and Vitale, S. and Vo, T. and Vocca, H. and Vorvick, C. and Vyatchanin, S. P. and Wade, A. R. and Wade, L. E. and Wade, M. and Wald, R. M. and Walet, R. and Walker, M. and Wallace, L. and Walsh, S. and Wang, G. and Wang, H. and Wang, J. Z. and Wang, W. H. and Wang, Y. F. and Ward, R. L. and Warden, Z. A. and Warner, J. and Was, M. and Watchi, J. and Weaver, B. and Wei, L.-W. and Weinert, M. and Weinstein, A. J. and Weiss, R. and Wellmann, F. and Wen, L. and Wessel, E. K. and We\ss{}els, P. and Westhouse, J. W. and Wette, K. and Whelan, J. T. and Whiting, B. F. and Whittle, C. and Wilken, D. M. and Williams, D. and Williamson, A. R. and Willis, J. L. and Willke, B. and Wimmer, M. H. and Winkler, W. and Wipf, C. C. and Wittel, H. and Woan, G. and Woehler, J. and Wofford, J. K. and Worden, J. and Wright, J. L. and Wu, D. S. and Wysocki, D. M. and Xiao, L. and Yamamoto, H. and Yancey, C. C. and Yang, L. and Yap, M. J. and Yazback, M. and Yeeles, D. W. and Yu, Hang and Yu, Haocun and Yuen, S. H. R. and Yvert, M. and Zadro\ifmmode \dot{z}\else \.{z}\fi{}ny, A. K. and Zanolin, M. and Zelenova, T. and Zendri, J.-P. and Zevin, M. and Zhang, J. and Zhang, L. and Zhang, T. and Zhao, C. and Zhou, M. and Zhou, Z. and Zhu, X. J. and Zimmerman, A. B. and Zucker, M. E. and Zweizig, J.},
  collaboration = {The LIGO Scientific Collaboration and the Virgo Collaboration},
  journal = {Phys. Rev. D},
  volume = {100},
  issue = {10},
  pages = {104036},
  numpages = {30},
  year = {2019},
  month = {Nov},
  publisher = {American Physical Society},
  doi = {10.1103/PhysRevD.100.104036},
  url = {https://link.aps.org/doi/10.1103/PhysRevD.100.104036}
}

@article{Alvarez-Gaume1983,
    author = "Alvarez-Gaume, Luis and Witten, Edward",
    editor = "Salam, A. and Sezgin, E.",
    title = "{Gravitational Anomalies}",
    reportNumber = "HUTP-83/A039",
    doi = "10.1016/0550-3213(84)90066-X",
    journal = "Nucl. Phys. B",
    volume = "234",
    pages = "269",
    year = "1984"
}

@article{Kanti2015,
   title={Gauss-Bonnet inflation},
   volume={92},
   ISSN={1550-2368},
   url={http://dx.doi.org/10.1103/PhysRevD.92.041302},
   DOI={10.1103/physrevd.92.041302},
   number={4},
   journal={Physical Review D},
   publisher={American Physical Society (APS)},
   author={Kanti, Panagiota and Gannouji, Radouane and Dadhich, Naresh},
   year={2015},
   month=aug }

@article{Yi2018,
    author = "Yi, Zhu and Gong, Yungui",
    title = "{Gauss\textendash{}Bonnet Inflation and the String Swampland}",
    eprint = "1811.01625",
    archivePrefix = "arXiv",
    primaryClass = "gr-qc",
    doi = "10.3390/universe5090200",
    journal = "Universe",
    volume = "5",
    number = "9",
    pages = "200",
    year = "2019"
}

@article{Chakraborty2018,
  title = {Inflation driven by Einstein-Gauss-Bonnet gravity},
  author = {Chakraborty, Sumanta and Paul, Tanmoy and SenGupta, Soumitra},
  journal = {Phys. Rev. D},
  volume = {98},
  issue = {8},
  pages = {083539},
  numpages = {15},
  year = {2018},
  month = {Oct},
  publisher = {American Physical Society},
  doi = {10.1103/PhysRevD.98.083539},
  url = {https://link.aps.org/doi/10.1103/PhysRevD.98.083539}
}

@article{Odintsov2018,
  title = {Viable inflation in scalar-Gauss-Bonnet gravity and reconstruction from observational indices},
  author = {Odintsov, S. D. and Oikonomou, V. K.},
  journal = {Phys. Rev. D},
  volume = {98},
  issue = {4},
  pages = {044039},
  numpages = {12},
  year = {2018},
  month = {Aug},
  publisher = {American Physical Society},
  doi = {10.1103/PhysRevD.98.044039},
  url = {https://link.aps.org/doi/10.1103/PhysRevD.98.044039}
}

@article{Rashidi2020,
   title={Gauss–Bonnet Inflation after Planck2018},
   volume={890},
   ISSN={1538-4357},
   url={http://dx.doi.org/10.3847/1538-4357/ab6a10},
   DOI={10.3847/1538-4357/ab6a10},
   number={1},
   journal={The Astrophysical Journal},
   publisher={American Astronomical Society},
   author={Rashidi, Narges and Nozari, Kourosh},
   year={2020},
   month=feb, pages={58} }

@article{Callan1985,
    author = "Callan, Jr., Curtis G. and Martinec, E. J. and Perry, M. J. and Friedan, D.",
    title = "{Strings in Background Fields}",
    reportNumber = "PRINT-85-0734 (PRINCETON)",
    doi = "10.1016/0550-3213(85)90506-1",
    journal = "Nucl. Phys. B",
    volume = "262",
    pages = "593--609",
    year = "1985"
}

@article{Gross1986,
    author = "Gross, David J. and Sloan, John H.",
    title = "{The Quartic Effective Action for the Heterotic String}",
    reportNumber = "NSF-ITP-87-02",
    doi = "10.1016/0550-3213(87)90465-2",
    journal = "Nucl. Phys. B",
    volume = "291",
    pages = "41--89",
    year = "1987"
}

@article{Metsaev1987,
  title={Order $\alpha$′ (two-loop) equivalence of the string equations of motion and the $\sigma$-model Weyl invariance conditions: Dependence on the dilaton and the antisymmetric tensor},
  author={R. R. Metsaev and Arkady A. Tseytlin},
  journal={Nuclear Physics},
  year={1987},
  volume={293},
  pages={385-419},
  url={https://api.semanticscholar.org/CorpusID:120615519}
}

@article{Hull1987,
    author = "Hull, C. M. and Townsend, P. K.",
    title = "{The Two Loop Beta Function for $\sigma$ Models With Torsion}",
    reportNumber = "PRINT-87-0074-REV, PRINT-87-0074",
    doi = "10.1016/0370-2693(87)91331-1",
    journal = "Phys. Lett. B",
    volume = "191",
    pages = "115--121",
    year = "1987"
}

@article{Sorbo2011,
   title={Parity violation in the Cosmic Microwave Background from a pseudoscalar inflaton},
   volume={2011},
   ISSN={1475-7516},
   url={http://dx.doi.org/10.1088/1475-7516/2011/06/003},
   DOI={10.1088/1475-7516/2011/06/003},
   number={06},
   journal={Journal of Cosmology and Astroparticle Physics},
   publisher={IOP Publishing},
   author={Sorbo, Lorenzo},
   year={2011},
   month=jun, pages={003–003} }

@article{Shiraishi2013,
   title={Parity violation in the CMB bispectrum by a rolling pseudoscalar},
   volume={2013},
   ISSN={1475-7516},
   url={http://dx.doi.org/10.1088/1475-7516/2013/11/051},
   DOI={10.1088/1475-7516/2013/11/051},
   number={11},
   journal={Journal of Cosmology and Astroparticle Physics},
   publisher={IOP Publishing},
   author={Shiraishi, Maresuke and Ricciardone, Angelo and Saga, Shohei},
   year={2013},
   month=nov, pages={051–051} }

@article{QUaD2009,
  title = {Parity Violation Constraints Using Cosmic Microwave Background Polarization Spectra from 2006 and 2007 Observations by the QUaD Polarimeter},
  author = {Wu, E. Y. S. and Ade, P. and Bock, J. and Bowden, M. and Brown, M. L. and Cahill, G. and Castro, P. G. and Church, S. and Culverhouse, T. and Friedman, R. B. and Ganga, K. and Gear, W. K. and Gupta, S. and Hinderks, J. and Kovac, J. and Lange, A. E. and Leitch, E. and Melhuish, S. J. and Memari, Y. and Murphy, J. A. and Orlando, A. and Piccirillo, L. and Pryke, C. and Rajguru, N. and Rusholme, B. and Schwarz, R. and O'Sullivan, C. and Taylor, A. N. and Thompson, K. L. and Turner, A. H. and Zemcov, M.},
  collaboration = {QUaD Collaboration},
  journal = {Phys. Rev. Lett.},
  volume = {102},
  issue = {16},
  pages = {161302},
  numpages = {4},
  year = {2009},
  month = {Apr},
  publisher = {American Physical Society},
  doi = {10.1103/PhysRevLett.102.161302},
  url = {https://link.aps.org/doi/10.1103/PhysRevLett.102.161302}
}

@article{Shiraishi2016,
   title={Parity violation in the CMB trispectrum from the scalar sector},
   volume={94},
   ISSN={2470-0029},
   url={http://dx.doi.org/10.1103/PhysRevD.94.083503},
   DOI={10.1103/physrevd.94.083503},
   number={8},
   journal={Physical Review D},
   publisher={American Physical Society (APS)},
   author={Shiraishi, Maresuke},
   year={2016},
   month=oct }

@article{Philcox2023,
  title = {Do the CMB Temperature Fluctuations Conserve Parity?},
  author = {Philcox, Oliver H. E.},
  journal = {Phys. Rev. Lett.},
  volume = {131},
  issue = {18},
  pages = {181001},
  numpages = {7},
  year = {2023},
  month = {Nov},
  publisher = {American Physical Society},
  doi = {10.1103/PhysRevLett.131.181001},
  url = {https://link.aps.org/doi/10.1103/PhysRevLett.131.181001}
}

@article{Mirshekari2012,
  title = {Constraining Lorentz-violating, modified dispersion relations with gravitational waves},
  author = {Mirshekari, Saeed and Yunes, Nicol\'as and Will, Clifford M.},
  journal = {Phys. Rev. D},
  volume = {85},
  issue = {2},
  pages = {024041},
  numpages = {12},
  year = {2012},
  month = {Jan},
  publisher = {American Physical Society},
  doi = {10.1103/PhysRevD.85.024041},
  url = {https://link.aps.org/doi/10.1103/PhysRevD.85.024041}
}

@article{Ezquiaga2021,
   title={Gravitational wave propagation beyond general relativity:  waveform distortions and echoes},
   volume={2021},
   ISSN={1475-7516},
   url={http://dx.doi.org/10.1088/1475-7516/2021/11/048},
   DOI={10.1088/1475-7516/2021/11/048},
   number={11},
   journal={Journal of Cosmology and Astroparticle Physics},
   publisher={IOP Publishing},
   author={Ezquiaga, Jose Maria and Hu, Wayne and Lagos, Macarena and Lin, Meng-Xiang},
   year={2021},
   month=nov, pages={048} }

@article{Ezquiaga2022,
   title={Modified gravitational wave propagation with higher modes and its degeneracies with lensing},
   volume={2022},
   ISSN={1475-7516},
   url={http://dx.doi.org/10.1088/1475-7516/2022/08/016},
   DOI={10.1088/1475-7516/2022/08/016},
   number={08},
   journal={Journal of Cosmology and Astroparticle Physics},
   publisher={IOP Publishing},
   author={Ezquiaga, Jose Maria and Hu, Wayne and Lagos, Macarena and Lin, Meng-Xiang and Xu, Fei},
   year={2022},
   month=aug, pages={016} }

@inproceedings{Shapiro1990GeneralRA,
  title={General Relativity and Gravitation, 1989: Solar system tests of general relativity: recent results and present plans},
  author={Irwin I. Shapiro},
  year={1990},
  url={https://api.semanticscholar.org/CorpusID:117428821}
}

@article{ReynaudJaekel, place={NL},
   title={Tests of general relativity in the Solar System},
   volume={168},
   ISSN={0074-784X},
   url={https://doi.org/10.3254/978-1-58603-990-5-203},
   DOI={10.3254/978-1-58603-990-5-203},
   abstractNote={Tests of gravity performed in the Solar System show a good agreement with general relativity. The latter is however challenged by observations at larger, galactic and cosmic scales which are presently cured by introducing &ldquo;dark matter&rdquo; or &ldquo;dark energy&rdquo;. A few measurements in the Solar System, particularly the so-called &ldquo;Pioneer anomaly&rdquo;, might also be pointing at a modification of gravity law at ranges of the order of the size of the Solar System. The present lecture notes discuss the current status of tests of general relativity in the Solar System. They describe metric extensions of general relativity which have the capability to preserve compatibility with existing gravity tests while opening free space for new phenomena. They present arguments for new mission designs and new space technologies as well as for having a new look on data of existing or future experiments.},
   number={Atom Optics and Space Physics},
   journal={Proceedings of the International School of Physics },
   publisher={IOS Press},
   author={Reynaud S. and Jaekel M.-T.},
   year={2009},
   pages={203–217} }

@article{Bergshoeff1989,
    author = "Bergshoeff, E. A. and de Roo, M.",
    title = "{The Quartic Effective Action of the Heterotic String and Supersymmetry}",
    reportNumber = "CERN-TH-5398-89",
    doi = "10.1016/0550-3213(89)90336-2",
    journal = "Nucl. Phys. B",
    volume = "328",
    pages = "439--468",
    year = "1989"
}

@article{Lovelock1970,
    author = "Lovelock, David",
    title = "{Divergence-free tensorial concomitants}",
    doi = "10.1007/BF01817753",
    journal = "Aequat. Math.",
    volume = "4",
    number = "1",
    pages = "127--138",
    year = "1970"
}

@article{Lovelock1971,
    author = "Lovelock, D.",
    title = "{The Einstein tensor and its generalizations}",
    doi = "10.1063/1.1665613",
    journal = "J. Math. Phys.",
    volume = "12",
    pages = "498--501",
    year = "1971"
}

@ARTICLE{Lanczos1932,
       author = {{Lanczos}, Cornel},
        title = "{Elektromagnetismus als nat{\"u}rliche Eigenschaft der Riemannschen Geometrie}",
      journal = {Zeitschrift fur Physik},
         year = 1932,
        month = mar,
       volume = {73},
       number = {3-4},
        pages = {147-168},
          doi = {10.1007/BF01351210},
       adsurl = {https://ui.adsabs.harvard.edu/abs/1932ZPhy...73..147L}
}

@article{Lanczos1938,
    author = "Lanczos, Cornelius",
    title = "{A Remarkable property of the Riemann-Christoffel tensor in four dimensions}",
    doi = "10.2307/1968467",
    journal = "Annals Math.",
    volume = "39",
    pages = "842--850",
    year = "1938"
}

\end{document}